\documentclass[twocolumn,prl,showpacs,preprintnumbers,amsmath,amssymb,superscriptaddress,bibliography]{revtex4-1}
\usepackage{graphicx}
\usepackage{setspace}

\usepackage{color}
\usepackage{ulem} 
\usepackage{enumerate}  
\graphicspath{{./figures/}}

\usepackage{ascmac}
\usepackage{amsmath,amsthm,amssymb,braket}
\usepackage{mathrsfs}

\usepackage[
colorlinks=true, citecolor=blue, urlcolor=blue, linkcolor=blue,
setpagesize=false, bookmarks=false, breaklinks=true
]{hyperref}
\usepackage{lineno}

\newcommand{\heta}{\hat{\eta}}

\newcommand{\boldeta}{\boldsymbol{\eta}}
\newcommand{\bsig}{\boldsymbol{\sigma}}

\newcommand{\br}{{\bf r}}

\newcommand{\ha}{\hat{a}}
\newcommand{\hc}{\hat{c}}

\newcommand{\hH}{\hat{H}}

\newcommand{\hn}{\hat{n}}
\newcommand{\hs}{\hat{s}}

\newcommand{\hU}{\hat{ U}}
\newcommand{\hV}{\hat{ V}}

\newcommand{\eqq}[1]{\begin{align} #1 \end{align}}

\begin{document}
\title{Spin, Charge and $\eta$-Spin Separation in One-Dimensional Photodoped Mott Insulators}

\author{Yuta Murakami}
\affiliation{Center for Emergent Matter Science, RIKEN, Wako, Saitama 351-0198, Japan}
\author{Shintaro Takayoshi}
\affiliation{Department of Physics, Konan University, Kobe 658-8501, Japan}
\author{Tatsuya Kaneko}
\affiliation{Department of Physics, Osaka University, Toyonaka, Osaka 560-0043, Japan}
\author{Andreas M. L\"{a}uchli}
\affiliation{Laboratory for Theoretical and Computational Physics, Paul Scherrer Institute, 5232 Villigen, Switzerland}
\affiliation{Institute of Physics, Ecole Polytechnique Fed\'{e}rale de Lausanne (EPFL), 1015 Lausanne, Switzerland}
\author{Philipp Werner}
\affiliation{Department of Physics, University of Fribourg, 1700 Fribourg, Switzerland}
\date{\today}

\begin{abstract} 
We show that effectively cold metastable states in one-dimensional photo-doped Mott insulators described by the extended Hubbard model exhibit spin, charge and $\eta$-spin separation.
Namely, their wave functions in the large on-site Coulomb interaction limit can be expressed as $|\Psi\rangle =|\Psi_{\rm charge}\rangle|\Psi_{\rm spin}\rangle |\Psi_{\rm \eta- spin}\rangle$,
which is analogous to the Ogata-Shiba states of the doped Hubbard model in equilibrium. 
Here, the $\eta$-spin represents the type of photo-generated pseudoparticle (doublon or holon). 
$|\Psi_{\rm charge}\rangle$ is determined by spinless free fermions, $|\Psi_{\rm spin}\rangle$ by the isotropic Heisenberg model in the squeezed spin space, and $|\Psi_{\rm \eta- spin}\rangle$ by the XXZ model in the squeezed $\eta$-spin space.
In particular, the metastable $\eta$-pairing and charge-density-wave (CDW) states correspond to the gapless and gapful states of the XXZ model.
The specific form of the wave function allows us to accurately determine the exponents of correlation functions. 
The form also suggests that the central charge of the $\eta$-pairing state is 3 and that of the CDW phase is 2, which we numerically confirm. Our study provides analytic and intuitive insights into the correlations between active degrees of freedom in photo-doped strongly correlated systems. 
\end{abstract}

\maketitle
{\it Introduction--}
Doping charge carriers into strongly correlated insulators provides a pathway to produce intriguing emergent phenomena such as high-$T_c$ superconductivity~\cite{Tokura_RMP,Dagotto1994RMP}.
In equilibrium, the doping concentration can be chemically controlled. An alternative nonequilibrium way of introducing charge carriers is {\it photo-doping}, where electrons are excited across the gap~\cite{Yonemitsu2008,Giannetti2016review,Basov2017review,Sentef2021nonthermal,Koshihara2022review}.  
The photo-doping of Mott insulators creates novel pseudoparticle excitations such as doublons and holons (in the single-band case), while the equilibrium system can host only one type of charge carrier.
Such additional degrees of freedom can lead to intriguing properties and nonthermal phases.
Important examples include photo-induced insulator-metal transitions~\cite{Iwai2003PRL,Okamoto2007PRL,Takahashi2008PRB,Oka2008PRB,Okamoto2010PRB,Eckstein2012,Ejima2022PRR} and charge density waves~\cite{Okamoto2014PRL,Mihailovic2014,Tohyama2012PRL}, and the control of magnetic~\cite{Afanasiev2019PRX,Jiajun2018Natcom} and superconducting orders~\cite{Rosch2008PRL,Suzuki2019,Wang2018PRL,Kaneko2019PRL,Werner2018PRB,Nikolaj2019JPSJ,Werner2019PRB,Ejima2020PRR}.

\begin{figure}[t]
  \centering
    \hspace{-0.cm}
    \vspace{0.0cm}
\includegraphics[width=85mm]{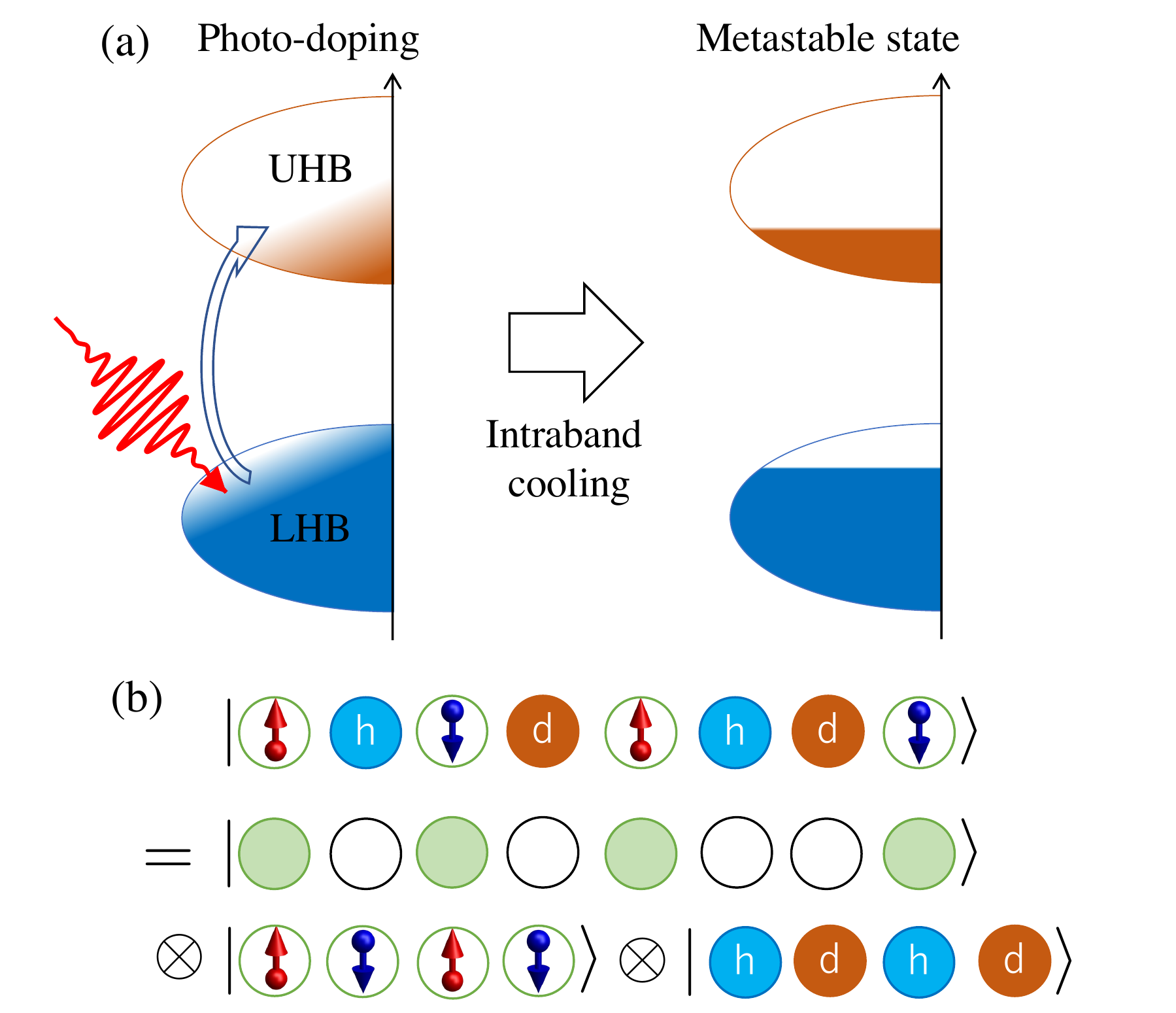} 
  \caption{(a) Schematic picture of the photo-doping and intraband cooling processes that result in a metastable state of the large-gap Mott insulator. UHB (LHB) stands for upper (lower) Hubbard band.
  (b) The wave function of the metastable state in the limit of $U\rightarrow \infty$ can be expressed as a direct product of the charge wave function, the spin wave function in the squeezed space and the $\eta$-spin wave function in the squeezed space. The green shaded circles in the charge wave function represent spinless fermions, while ``h" and ``d" stand for a holon and a doublon, respectively. }
  \label{fig:schematic}
\end{figure}

In systems with a large Mott gap, the life-time of the photo-doped pseudoparticles becomes exponentially enhanced~\cite{Strohmaier2010PRL,Zala2013PRL,Mitrano2014,Sensarma2010PRB,Martin2011PRB,Zala2014PRB,Nevola2021PRB}.
In such a situation, an intraband cooling of the photo-doped pseudoparticles may occur, while their density remains approximately constant. 
This results in a metastable steady state (a pseudoequilibrium state)~\cite{Takahashi2002PRB_A,Takahashi2002PRB,Takahashi2005PRB,Rosch2008PRL,Ishihara2011PRL,Jiajun2020PRB,Jiajun2021PRB,Murakami2022,Yokoyama2022arxiv}, analogous to the case of photo-doped semiconductors~\cite{haug_quantum_1990,Keldysh1986review,Asano2014JPSJ}, see Fig.~\ref{fig:schematic}(a).
It has been shown that such metastable states can host unique phases such as $\eta$-pairing~\cite{Jiajun2020PRB,Murakami2022}, chiral superconducting phases \cite{Jiajun2022arxiv}, and exotic spin/orbital orders~\cite{Jiajun2018Natcom,Jiajun2021PRB,Werner2021PRBL}.
Since different types of charge carriers are present in photo-doped systems, it is crucial to understand the correlations between the active degrees of freedoms.
However, the metastable states of photo-doped strongly correlated systems have been mainly studied numerically so far~\cite{Takahashi2002PRB_A,Takahashi2002PRB,Takahashi2005PRB,Ishihara2011PRL,Ishihara2012PRB,Jiajun2020PRB,Jiajun2021PRB,Murakami2022,Yokoyama2022arxiv}, and analytical or intuitive insights are limited.

Here we reveal 
the nature of the metastable states and the correlations between the active degrees of freedom in photo-doped one-dimensional Mott insulators.
We show that the wave functions of the metastable states in the limit of large on-site Coulomb interaction exhibit spin, charge and $\eta$-spin separation, see Fig.~\ref{fig:schematic}(b).
The $\eta$-spin represents the type of pseudoparticle: doubly occupied site (doublon) or empty site (holon).
Our results provide a comprehensive understanding of the character of the photo-induced metastable phases in one-dimensional systems and reveal the similarities and differences between photo-doped and chemically-doped systems.

{\it Results--} We focus on the one-dimensional extended Hubbard model,  
\eqq{
\hat{H} =-t_{\rm hop}\sum_{i,\sigma} (\hc^\dagger_{i,\sigma}\hc_{i+1,\sigma} +h.c.) + \hH_U + \hH_V, \label{eq:Hamiltonian}
}
and assume that electrons are excited across the Mott gap via photo-excitation.
Similar setups can be considered with cold atoms~\cite{Rosch2008PRL}.
$\hH_U=U \sum_i (\hn_{i\uparrow}-\frac{1}{2}) ( \hn_{i\downarrow}-\frac{1}{2}) $ is the on-site interaction and $ \hH_V \nonumber=V \sum_{i} (\hn_i-1) (\hn_{i+1}-1)$ is the nearest-neighbor interaction. $\hc^\dagger_{i\sigma}$ is the creation operator of a fermion with spin $\sigma$ at site $i$, 
$\hn_{i\sigma} =\hc^\dagger_{i\sigma} \hc_{i\sigma}$, $\hn_{i} = \hn_{i\uparrow} + \hn_{i\downarrow}$, and $t_{\rm hop}$ is the hopping parameter.
When the Mott gap is large enough, the recombination time of the created doublons and holons becomes exponentially long~\cite{Strohmaier2010PRL,Zala2013PRL,Mitrano2014,Sensarma2010PRB,Martin2011PRB,Zala2014PRB}.
Thus, intraband relaxation due to scattering events and coupling to the environment is expected to bring the system into a (intraband thermalized) steady state with a fixed number of doublons and holons, see Fig.~\ref{fig:schematic}(a).
As previously discussed, such a quasi-steady state can be described with the effective Hamiltonian obtained by a Schrieffer-Wolff transformation~\cite{MacDonald1988PRB} from the original Hamiltonian~\eqref{eq:Hamiltonian}~\cite{Takahashi2002PRB_A,Takahashi2002PRB,Takahashi2005PRB,Rosch2008PRL,Ishihara2011PRL,Ishihara2012PRB,Jiajun2020PRB,Jiajun2021PRB,Murakami2022}, see also the Supplemental Material (SM)~\cite{SM}. This effective 
Hamiltonian explicitly conserves the number of doublons and holons. Up to $\mathcal{O}(t^2_{\rm hop}/U)$, it takes the form
\eqq{
 \hH_{\rm eff}  =&   \hH_U + \hH_{\rm kin} + \hH_V \nonumber\\
 &                          +  \hH_{\rm spin,ex} +  \hH_{\rm dh,ex} + \hH_{U,\rm{shift}} + \hH_{\rm 3-site}, \label{eq:Heff}
}
where  $\hH_{\rm kin} = -t_{\rm hop}\sum_{\langle i,j\rangle,\sigma} \hat{\bar{n}}_{i,\bar{\sigma}} (\hc^\dagger_{i,\sigma}\hc_{j,\sigma} + h.c.)  \hat{\bar{n}}_{j,\bar{\sigma}} -t_{\rm hop}\sum_{\langle i,j\rangle,\sigma} \hn_{i,\bar{\sigma}} (\hc^\dagger_{i,\sigma}\hc_{j,\sigma} +h.c.)  \hn_{j,\bar{\sigma}}$ represents the hopping of a doublon or a holon, $\bar{\sigma}$ is the opposite spin of $\sigma$, and $\hat{\bar{n}}_{i,\sigma}=1- \hn_{i,\sigma}$.
The other terms are proportional to $J_{\rm ex}\equiv\frac{4t^2_{\rm hop}}{U}$. $\hH_{\rm spin,ex}= J_{\rm ex} \sum_{\langle i,j\rangle} \hat{\bf  s}_i\cdot \hat{\bf  s}_{j}$
is the spin exchange term, and $\hH_{\rm dh,ex}  = -J_{\rm ex}\sum_{\langle i,j\rangle} [ \heta^x_i \heta^x_{j} + \heta^y_i \heta^y_{j} + \heta^z_i \heta^z_{j}]$ is the exchange term for doublons and holons on neighboring sites. Here the spin operators are $\hat{{\bf s}} =\frac{1}{2}\sum_{\alpha,\beta=\uparrow,\downarrow} \hc^\dagger_\alpha \boldsymbol{\sigma}_{\alpha\beta} \hc_\beta$ with $\boldsymbol{\sigma}$ denoting the Pauli matrices, and we introduced the $\eta$-spin operators as $\heta^+_i = (-)^i\hc^\dagger_{i\downarrow}  \hc^\dagger_{i\uparrow}$, $\heta^-_i = (-)^i\hc_{i\uparrow}  \hc_{i\downarrow}$ and $\heta^z_i =\frac{1}{2}(\hn_i-1)$~\cite{Yang1989PRL,Essler2005,Nakagawa2022arxiv}.
$\hH_{U,\rm{shift}}$ describes the shift of the local interaction and $\hH_{\rm 3-site}$ represents three-site terms such as correlated doublon hoppings, see SM~\cite{SM}.
In equilibrium (without doublons), the model corresponds to the $t$-$J$ model when $\hH_{\rm 3-site}$ is neglected \footnote{The $t$-$J$ model using $\eta$-spins was also introduced for the attractive Hubbard model~\cite{Rojo1990PRB}.}.
In the following, we denote the model without $\hH_{\rm 3-site}$ by $\hH_{\rm eff2}$.
When $V=0$, $\hat{H}$, $ \hH_{\rm eff}$ and $\hH_{\rm eff2}$ host an SU$(2)$ symmetry of the doublon-holon sector~\cite{Yang1989PRL,Zhang1990PRL} that corresponds to the spin SU$(2)$ symmetry via a particle-hole (Shiba) transformation \cite{Shiba1972}.

We consider an effectively cold system with arbitrary filling,
whose state is described by the ground state of the effective Hamiltonian for a given number of doublons and holons,
i.e., we assume that the system is thermalized into the lowest energy state for the given constraint.
 We show that the corresponding wave function can be expressed as the direct product of the charge, spin and $\eta$-spin wave functions in the limit of $J_{\rm ex} \rightarrow 0$ with $V/J_{\rm ex}={\rm const} $,
similar to the Ogata-Shiba state of the doped Hubbard model in equilibrium~\cite{Ogata1990PRB,Shiba1991}.
 To be more specific, we set the system size to $L$ and the number of singly occupied sites to $N_s$, so that the number of doublons and holons (the number of $\eta$ spins) is  $N_{\eta}=L-N_s$.
Now we introduce the Hilbert space 
 \eqq{
 \mathcal{H}' &= \Bigl\{ |\br\rangle|\bsig\rangle|\boldeta\rangle\equiv \Big(\prod_{r\in \br} \hc^\dagger_{r}\Big)|{\rm vac}\rangle|\bsig\rangle|\boldeta\rangle \nonumber \\
  & \;\;\;\;\;\; \text{: $\# \br =\# \bsig= N_s$ and $\# \boldeta = N_{\eta}$} \Bigl\}.
 }
 Here $\br$, $\bsig$ and $\boldeta$ are sets of space, spin and $\eta$-spin indices, $\hc^\dagger_r$ is a creation operator of a spin-less fermion (SF), and $\#$ indicates the number of elements. $\eta$ takes the values $\uparrow$ or $\downarrow$ and $\br = \{r_{N_s}, \cdots, r_1\}$ with $L\geq r_{N_s}>r_{N_s-1}>\cdots > r_{1}\geq 1$. 
 We identify this Hilbert space with the original Hilbert space using the unitary transformation $\hU:  \mathcal{H} \rightarrow  \mathcal{H}'$ defined by
 \eqq{
 \hU \Bigg( \Big(\prod_{i=1}^{N_s} \hc^\dagger_{r_i,\sigma_i}\Big) \Big(\prod_{j=1}^{N_{\eta}} \ha^\dagger_{\bar{r}_j,\eta_j}\Big)|{\rm vac}\rangle  \Bigg) = \Big(\prod_{r\in \br} \hc^\dagger_{r}\Big)|{\rm vac}\rangle|\bsig\rangle|\boldeta\rangle.
 }
 Here, $\bar{\br} = \{\bar{r}_{N_{\eta}}, \cdots \bar{r}_1\}$ with $L\geq \bar{r}_{N_{\eta}}>\bar{r}_{N_{\eta}-1}>\cdots > \bar{r}_{1}\geq 1$, $\br \cup \bar{\br} = \{L,L-1,\cdots,1\}$, $\ha^\dagger_{\bar{r},\uparrow} = (-)^{\bar{r}} \hc^\dagger_{\bar{r}\downarrow} \hc^\dagger_{\bar{r}\uparrow}$ and $\ha^\dagger_{\bar{r},\downarrow} = 1$. With this identification, $|{\bf r}\rangle$ is the basis of SF and $|\bsig\rangle$ ($|\boldeta\rangle$) is the basis of the squeezed spin ($\eta$-spin) space. 
 Note that the $\eta$-spin configuration represents the sequence of doublons and holons.
 As shown below, Hamiltonians ruling the $\sigma$ and $\eta$ spaces are not fully symmetric due to the staggering of the doublons.

The wave function in the limit of $J_{\rm ex} \rightarrow 0$  can be constructed by degenerate perturbation theory~\cite{Shiba1991}.
For $J_{\rm ex} = V=0$, the eigenstates of $ \hH_{\rm eff}$ are degenerate with respect to the configurations of spins and $\eta$-spins.
This is because $\hH_{\rm kin}$ never exchanges the positions of spins or those of doublons and holons.
Specifically, one can show that   
 $ \hU \hH_{\rm kin}  \hU^\dagger  = -t_{\rm hop}\sum_{\langle i,j\rangle} (\hc^\dagger_{i}\hc_{j} + \text{h.c.})$ $(\equiv \hH_{0,{\rm SF}})$.
 This means that in the representation of $\mathcal{H}'$ the ground state for $J_{\rm ex} = V=0$ 
 can be described as $|\Psi^{\rm GS}_{\rm SF}\rangle|\Psi_{\sigma,\eta}\rangle$, where $|\Psi^{\rm GS}_{\rm SF}\rangle$ is the ground state of $ \hH_{0,{\rm SF}}$ and  
 $|\Psi_{\sigma,\eta}\rangle$ is an arbitrary spin and $\eta$-spin wave function.
 The remaining spin/$\eta$-spin degeneracy of $2^{N_s}\cdot 2^{N_{\eta}}$ is lifted by the terms of $\mathcal{O}(J_{\rm ex})$.
 Within lowest-order degenerate perturbation theory, the wave function of the spin and $\eta$-spin is obtained by 
 the $\mathcal{O}(J_{\rm ex})$ terms projected to $|\Psi^{\rm GS}_{\rm SF}\rangle|\bsig\rangle|\boldeta\rangle$.
 In the resultant projected Hamiltonian, the squeezed spin and $\eta$-spin spaces are decoupled, and the corresponding Hamiltonians become
(SQ stands for squeezed space)
 \eqq{
 \hH_{\rm spin}^{\rm (SQ)} &= J^s_{\rm ex} \sum_i  \hat{\bf  s}_{i+1}\cdot \hat{\bf  s}_{i} ,  \nonumber\\
  \hH_{\rm \eta-spin}^{\rm (SQ)} &= -J^{\eta}_{X} \sum_j (\hat{\eta}^x_{j+1}\hat{\eta}^x_{j} + \hat{\eta}^y_{j+1}\hat{\eta}^y_{j}) +  J^{\eta}_Z\sum_j \hat{\eta}^z_{j+1}\hat{\eta}^z_{j},\nonumber
 }
 with $J^{\rm s}_{\rm ex}=(\tilde{x}-\tilde{x}') J_{\rm ex} $, $J_X^{\eta} = (\tilde{y}-\tilde{y}') J_{\rm ex}$ and  $J_Z^\eta = -(\tilde{y}-\tilde{y}') J_{\rm ex} + 4 \tilde{y} V$. 
Here $\tilde{x},\tilde{x}', \tilde{y}$ and $\tilde{y}'$ are the renormalization factors determined by $|\Psi^{\rm GS}_{\rm SF}\rangle$.
With $n_s = N_s/L$ and $n_{\eta} = N_{\eta}/L$ and in the limit $L\rightarrow \infty$ they can be expressed as 
  \eqq{
& \tilde{x} = n_s - \frac{\sin^2(\pi n_s)}{\pi^2 n_s},\;\tilde{x}' = \frac{\sin(2\pi n_s)}{2\pi} - \frac{\sin^2(\pi n_s)}{\pi^2 n_s}, \nonumber \\
& \tilde{y} = n_{\eta} - \frac{\sin^2(\pi n_{\eta})}{\pi^2 n_{\eta}},\;\tilde{y}' = \frac{\sin(2\pi n_{\eta})}{2\pi} - \frac{\sin^2(\pi n_{\eta})}{\pi^2 n_{\eta}}.\nonumber
 }
 Here $ \tilde{x}$ and $\tilde{y}$ are the contributions from the 2-site terms of  $\mathcal{O}(J_{\rm ex})$, while $ \tilde{x}'$ and $\tilde{y}'$ are those from the 3-site terms.
Note that $\hH_{\rm \eta-spin}^{\rm (SQ)}$ becomes the ferromagnetic Heisenberg model ($J^{\eta}_{X}=- J^{\eta}_Z>0$) for $V=0$.
 Thus, the wave function (in $\mathcal{H}'$) takes the form 
 \eqq{
  |\Psi\rangle = |\Psi^{\rm GS}_{\rm SF}\rangle|\Psi^{\rm GS}_{\sigma}\rangle|\Psi^{\rm GS}_{\eta}\rangle, \label{eq:WF}
 } where 
$ |\Psi^{\rm GS}_{\sigma}\rangle$ is the ground state of $\hH_{\rm spin}^{\rm (SQ)}$ and  $|\Psi^{\rm GS}_{\eta}\rangle$ is that of $\hH_{\rm \eta-spin}^{\rm (SQ)}$.
For more details, see SM~\cite{SM}.
The form of $|\Psi^{\rm GS}_{\rm SF}\rangle$ and $|\Psi^{\rm GS}_{\sigma}\rangle$ is independent of the ratio of doublons and holons, and, in particular, these states are the same as those in the equilibrium doped Hubbard model at the doping level $n_{\rm holes}=n_\eta$~\cite{Ogata1990PRB,Shiba1991}. 
This implies that the effects of photo-doping and chemical doping on the spins are essentially the same, which is consistent with previous numerical analyses~\cite{Werner2012PRB, Eckstein2014PRL,Murakami2022}.

 \begin{figure}[t]
  \centering
    \hspace{-0.cm}
    \vspace{0.0cm}
\includegraphics[width=65mm]{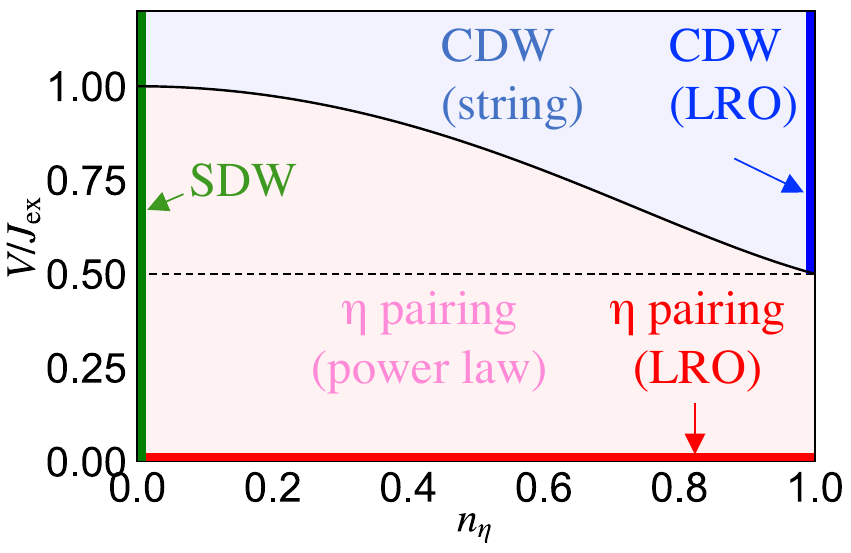} 
  \caption{Phase diagram of the photo-doped one-dimensional Mott insulator described by $\hH_{\rm eff}$ in the limit $J_{\rm ex}\rightarrow 0$. The phase boundary (black solid line) corresponds to an SU$(2)$ symmetric point of $\hH_{\rm \eta-spin}^{\rm (SQ)}$, i.e. $V/J_{\rm ex} = \frac{\tilde{y}-\tilde{y}'}{2\tilde{y}}$.  The horizontal dashed line indicates the phase boundary for the system described by $\hH_{\rm eff2}$.  }
  \label{fig:phase_diagram}
\end{figure}

Now we focus on half filling and discuss the implications of the exact form of the wave function for the origin of the different phases.
The $\eta$-spin sector hosts the phase transition between the gapless and gapful phases of the XXZ model, which is controlled by the ratio between $J_{\rm ex}$ and $V$.
As seen below, these states are characterized by the behavior of the correlation functions of the $\eta$-spins, i.e. $\chi_{\rm \eta,a}(r) \equiv  \langle \heta^a(r) \heta^a(0)\rangle$.
 Namely, the gapless phase corresponds to the $\eta$-pairing phase, where the pair correlation  $\chi_{\rm \eta -pair}\equiv \chi_{\eta,x}$ is dominant. On the other hand, the gapful phase 
 corresponds to the CDW phase, where the charge correlation $\chi_{\rm charge}\equiv \chi_{\eta,z}$ is dominant. 
 True long-range order (LRO) is realized at $V=0$ for the $\eta$-pairing phase \footnote{At $V=0$, the order parameter $\hat{\eta}^+\equiv \sum_i \hat{\eta}^+_i$ commutes with the Hamiltonian and we consider a ground state for a given number of doublons and holons.},
  while a LRO CDW is realized at $n_\eta=1$ and $V>\frac{J_{\rm ex}}{2}$.
 Apart from these limits, we have quasi-long-range orders (power law decay of correlations). 
Note that the appearance of $\eta$-pairing in nonthermal states has been recently discussed~\cite{Kaneko2019PRL,Jaksch2019PRL,Ejima2020PRR,Jiajun2020PRB,Peronaci2020PRB,Tindall2020PRL} in relation with the photo-induced superconducting-like phases~\cite{Cavalleri2011,Mitrano2016,Cavalleri2018review,Suzuki2019,Buzzi2020PRX}.
Furthermore, we emphasize that LRO is realized in the squeezed $\eta$-spin space for the CDW phase, which is reminiscent of the string order in the Haldane phase~\cite{Marcel1989PRB}.
The phase transition occurs at the SU$(2)$ point of $\hH_{\rm \eta-spin}^{\rm (SQ)}$ ($J^{\eta}_{X}= J^{\eta}_Z>0$), see Fig.~\ref{fig:phase_diagram}. 
 For $\hH_{\rm eff2}$ (without $\hH_{\rm 3-site}$), $\Delta(\equiv J^\eta_Z/J^\eta_X)$ and thus the phase boundary are independent of the filling,
 which consistently explains a previous numerical result ~\cite{Murakami2022}.
 On the other hand, for $\hH_{\rm eff}$, the ratio $\Delta$ depends on the filling due to the effects of the 3-site term $\tilde{y}'$. 
 In particular, the 3-site term is found to favor the $\eta$-pairing phase.

 \begin{figure}[t]
  \centering
    \hspace{-0.cm}
    \vspace{0.0cm}
\includegraphics[width=89mm]{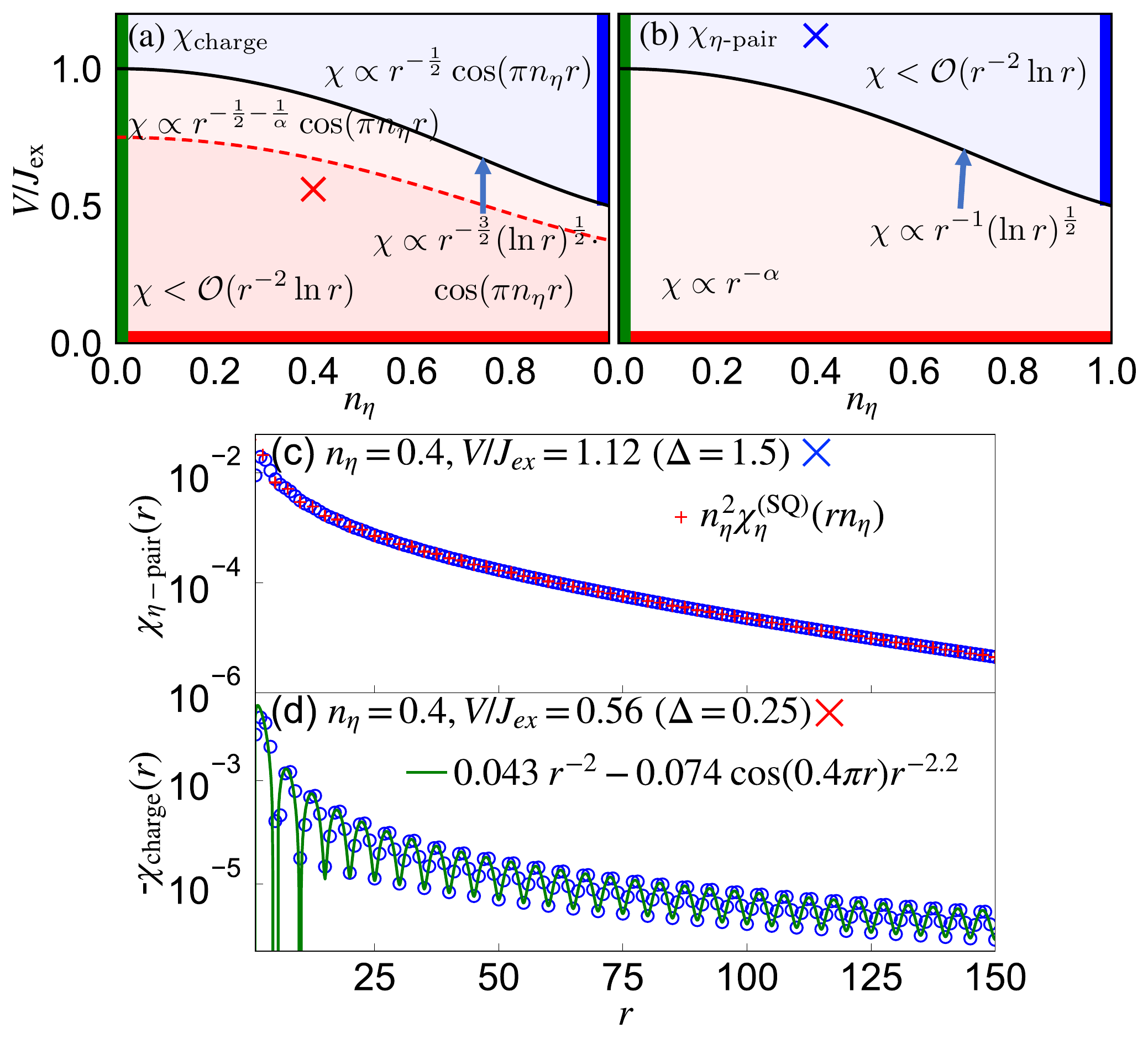} 
  \caption{(a)(b) Asymptotic behavior analytically obtained for (a) $\chi_{\rm charge}$ and (b)  $\chi_{\rm \eta -pair}$. 
  The dashed line at  $\Delta=\frac{1}{2}$ in (a) is the boundary of different expressions.
The thick green line corresponds to the SDW, the thick blue line to the CDW with LRO and the thick red line to the $\eta$-pairing phase with LRO.
  (c)(d) Numerically evaluated correlation functions for (c) the $\eta$-pairing phase and (d) the CDW phase using Eq.~\eqref{eq:correlations} and the iTEBD results for the XXZ model (blue circles).
  The corresponding points are indicated with crosses in panels (a)(b). We also show the correlations estimated by the conjecture $\chi_{\rm \eta -pair}\simeq n_\eta^2\chi^{\rm (SQ)}_{\eta}(rn_{\eta})$ as well as the fit with $C_1 r^{-2} + C_2 r^{-\frac{1}{2}-\frac{1}{\alpha}}\cos(\pi n_{\eta}r)$.}
  \label{fig:correlation}
\end{figure}

The exact form of the wave function allows us to evaluate the asymptotic behavior of the correlation functions analytically or numerically. Here we extend the analyses for spin correlations of the equilibrium Hubbard model~\cite{Sorella1990PRL,Pruschke1991PRB}.
Since the spin correlations of the metastable state are the same as those for the equilibrium Hubbard model, i.e. $\langle \hs^a(r) \hs^a(0)\rangle\propto  \cos(\pi n_s r)r^{-\frac{3}{2}} (\ln r)^\frac{1}{2}$, we focus on the $\eta$-spin correlation functions $\chi_{\eta,a}(r)$. Note that despite the apparent similarity between the squeezed spin and $\eta$ space there are crucial differences in
the pairing correlations.
Using expression~\eqref{eq:WF}, the correlation functions are expressed as 
\eqq{
 \chi_{\eta,a}(r) = \sum_{m=2}^{r+1} \bar{Q}^r_{\rm SF}(m) \chi^{\rm (SQ)}_{\eta,a} (m-1). \label{eq:correlations}
}
Here  $\bar{Q}^r_{\rm SF}(m) = \langle \hat{\bar{n}}_0 \hat{\bar{n}}_r \delta(\sum_{l=0}^r \hat{\bar{n}}_l -m) \rangle_{\rm SF} $,  which is determined by $|\Psi^{\rm GS}_{\rm SF}\rangle$, 
  is the probability that the system has $m$ doublons or holons in $[0,r]$.
$\chi^{\rm (SQ)}_{\eta,a}(m) = \langle \heta^a(m) \heta^a(0) \rangle_{\rm \eta-spin,squeezed}$
 is the correlation function in the squeezed $\eta$-spin space.
  Numerically,  $\bar{Q}^r_{\rm SF}(m)$ and  $\chi^{\rm (SQ)}_{\eta,a}(m)$ can be efficiently evaluated in the thermodynamic limit.
We use the expression for the Fourier components and perform an inverse Fourier transformation to obtain $\bar{Q}^r_{\rm SF}(m)$ ~\cite{Pruschke1991PRB, SM}, 
  while  the infinite time-evolving block decimation (iTEBD)~\cite{Vidal2003PRL} for the XXZ model can be used to calculate $\chi^{\rm (SQ)}_{\eta,a}(m)$.
Moreover, we can also gain analytic insights using the knowledge of the asymptotic behavior of the correlation functions of the XXZ model~\cite{Giamarchi_book,Lukyanov99PRB,Lukyanov03NPB}
  and the moments of $\bar{Q}^r_{\rm SF}(m)$ up to the second one~\cite{Sorella1990PRL}.
The former can be expressed with $\alpha\equiv 1 - \frac{1}{\pi}\arccos(\Delta)$, which is a control parameter of the Tomonaga-Luttinger liquid, and the latter indicates that most of the weight of $\bar{Q}^r_{\rm SF}(m)$ is at $rn_{\eta}$.
  From these facts, if the asymptotic form of  $\chi^{\rm (SQ)}_{\eta}(m)$ is $(-)^m f(m)$ with $f(m)$ being a smooth function, one can prove that 
  \eqq{
&  \sum_{m=2}^{r+1} \bar{Q}^r_{\rm SF}(m) (-)^m f(m)  \simeq  \Bigg\{ \sum_{m=2}^{r+1} \bar{Q}^r_{\rm SF}(m) (-)^m\Bigg\} f(\overline{\langle m\rangle} ). \label{eq:up_limit}
  } 
 Here $\overline{\langle m\rangle} = n_{\eta} r +1$. If $\chi^{\rm (SQ)}_{\eta}(m)\simeq f(m)$, the equation without $(-)^m$ is satisfied. 
 See SM for the detailed meaning of the equality $\simeq$ and the derivation.
 Since we have  $\{ \sum_{m=2}^{r+1} \bar{Q}^r_{\rm SF}(m) (-)^m\} \propto \frac{\cos(\pi n_{\eta} r)}{r^{\frac{1}{2}}}$~\cite{Sorella1990PRL,Holger1990PRB} and $\sum_{m=2}^{r+1} \bar{Q}^r_{\rm SF}(m)=n_{\eta}^2$ (in leading order in $r$),  one can obtain the asymptotic form of the correlation functions.
 Equation~\eqref{eq:up_limit} shows that the decay of $\eta$-spin correlations in real space originates from that in the squeezed space and the contribution from the intercalated singly-occupied sites.
 The latter is determined by $|\Psi^{\rm GS}_{\rm SF}\rangle$, and has a different impact depending on whether the correlation functions in the squeezed space are staggered or not. In particular, the pairing correlation is not affected by the SF background, while the charge correlations can be affected like the spin correlations.
 
The asymptotic forms obtained analytically for $\chi_{\rm charge}$ and $\chi_{\rm \eta-pair}$ are summarized in Figs.~\ref{fig:correlation}(a)(b).
The magnitude relation of the exponents of these correlation functions changes at the SU$(2)$ point of $\hH_{\rm \eta-spin}^{\rm (SQ)}$. 
Note that this SU$(2)$ symmetry is an emergent symmetry in the squeezed space, which is absent in the original Hamiltonian.
For $0<n_\eta<1$, $\chi_{\rm charge}$ shows an exponent of $1/2$ in the CDW phase due to the contribution from the SF part, although it shows LRO in the squeezed space.
On the other hand, the analytic argument based on Eq.~\eqref{eq:up_limit} does not allow to make exact statements for the components decaying faster than $\mathcal{O}(\frac{\ln r}{r^2})$.
To analyze this point, we numerically evaluate the correlation functions, see  Figs.~\ref{fig:correlation}(c)(d).
Firstly, our results verify the conjecture $\chi_{\eta-\rm{pair}}(r)\simeq n_{\eta}^2\chi^{\rm (SQ)}_{\eta,x}(rn_{\eta})$ and its applicability even in the CDW regime, where $\chi_{\eta-\rm{pair}}(r)$ decays exponentially, see Fig.~\ref{fig:correlation}(c).
Secondly, Fig.~\ref{fig:correlation}(d) shows that Eq.~\eqref{eq:up_limit} is practically applicable for the leading and the sub-leading terms of $\chi_{\rm charge}$ decaying faster than $\mathcal{O}(\frac{\ln r}{r^2})$, i.e. $\chi_{\rm charge}(r)\simeq C_1 r^{-2} + C_2 r^{-\frac{1}{2}-\frac{1}{\alpha}}\cos(\pi n_{\eta}r)$.

 \begin{figure}[t]
  \centering
    \hspace{-0.cm}
    \vspace{0.0cm}
\includegraphics[width=80mm]{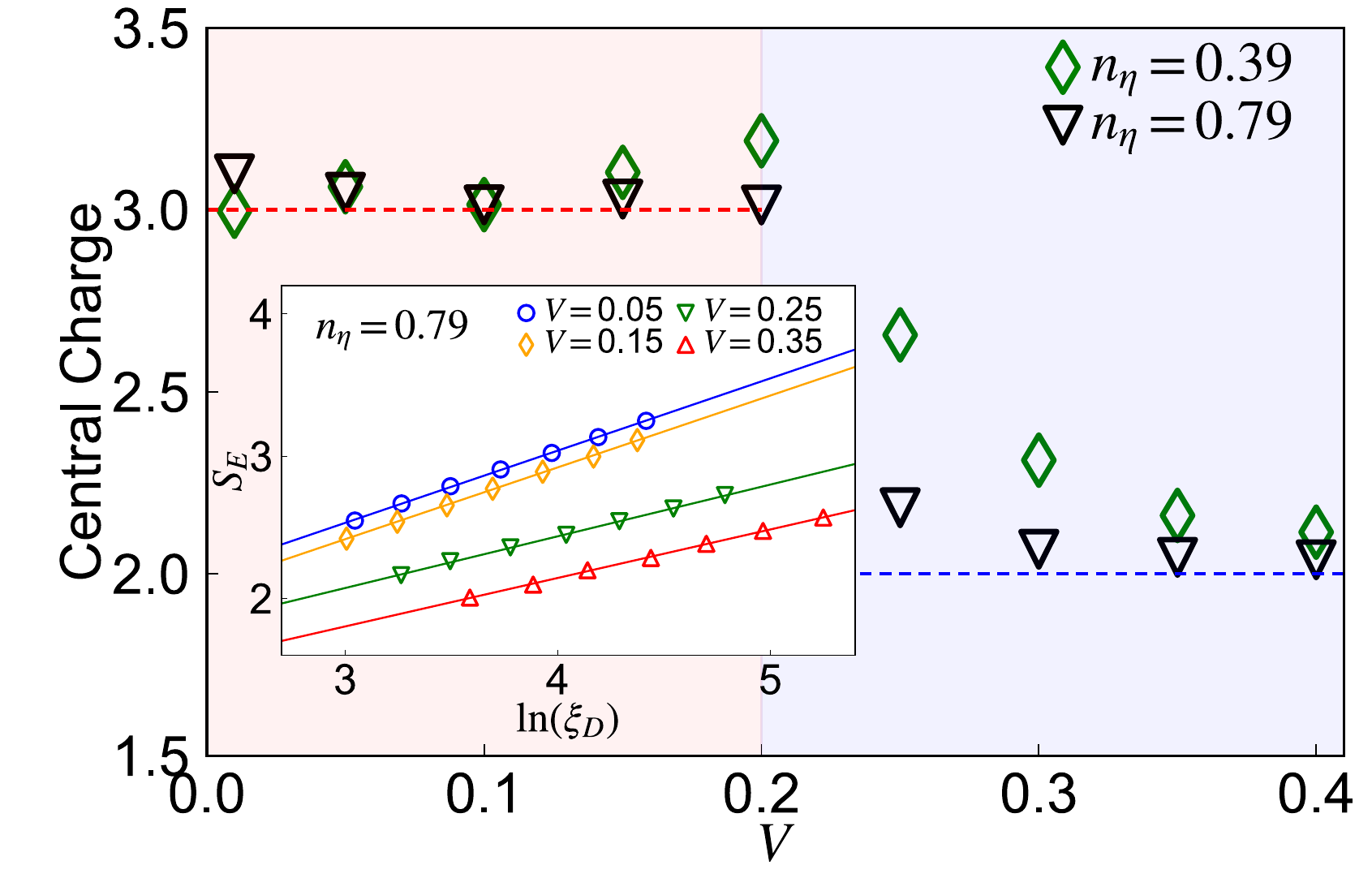} 
  \caption{Central charge of the ground state of $\hH_{\rm eff2}$ for $J_{\rm ex} = 0.4$ and the indicated values of $n_{\eta}$. To evaluate the central charge, we apply Eq.~\eqref{eq:cc_scaling} to the iTEBD results with $D\in [200,1600]$.
  The red shaded area indicates the stability region of the $\eta$-pairing phase for $J_{\rm ex} \rightarrow 0$, while the blue shaded area shows that of the CDW phase.
  The inset plots the relation between $S_E$ and $\xi_D$ and the corresponding linear fits. }
  \label{fig:central_charge}
\end{figure}

The expression \eqref{eq:WF} also provides valuable insights into the physical nature of the metastable phases.
One important quantity that characterizes one-dimensional systems is the central charge ($c$), 
which counts to the number of gapless degrees of freedoms~\cite{Giamarchi_book}.
In equilibrium, the doped Hubbard model exhibits $c=2$, because of the massless modes both in the spin and charge sectors~\cite{Ogata1991PRL,Essler2013PRL}.
On the other hand, the exact form of the wave-function \eqref{eq:WF} suggests that the metastable state possesses three degrees of freedoms.
The wave functions of the charge and spin sectors are those of gapless states (i.e. doped free fermions and the isotropic Heisenberg model), while that of the $\eta$-spin sector 
corresponds to the gapless state or the gapful state of the $\eta$-XXZ model for the $\eta$-pairing state and the CDW state, respectively.
Thus, one naturally expects that $c=3$ in the $\eta$-pairing state and $c=2$ in the CDW state.
To confirm this, we perform iTEBD simulations on the effective model $\hH_{\rm eff2}$ for various cut-off dimension ($D$) and extract $c$ from the relation~\cite{Pollmann2013PRB} 
\eqq{
S_E = \frac{c}{6} \ln(\xi_D) + s_0 \label{eq:cc_scaling}.
}
Here $S_E$ is the entanglement entropy, $s_0$ is a constant and $\xi_D$ is the correlation length evaluated from the second-largest eigenvalue of the transfer matrix, see SM.
In Fig.~\ref{fig:central_charge}, we show the central charge for $\hH_{\rm eff2}$ with $J_{\rm ex}=0.4$, which is extracted using Eq.~\eqref{eq:cc_scaling} and the linear fit of the iTEBD results [see the inset of Fig.~\ref{fig:central_charge}].
The results indeed confirm the above expectation.
We emphasize that the emergence of a $c=3$ state in the Hubbard model is hardly expected in equilibrium and reflects the metastable nature of the state.

{\it Conclusion--}
We showed that the additional degrees of freedom activated by photo-doping lead to peculiar types of quantum liquids absent in equilibrium.
In particular, we revealed the intriguing structure of the correlations between active degrees of freedom in photo-doped one-dimensional strongly correlated systems, i.e. the spin-charge-$\eta$-spin separation.
Our results open a new avenue for studying metastable states in one-dimensional systems and raise interesting questions.
Firstly, in contrast to the equilibrium Hubbard model, the weak coupling regime is not well-defined, and the relation between the lattice model and the corresponding conformal field theory is not clear.
Construction of the field theory for the metastable states is an important future task.
Secondly, we provide a rigorous basis for the future development of a bosonization approach~\cite{Furusaki2007PRL,Furusaki2007PRB}.
With such an approach, one can better understand the spectral features of the photo-doped systems and the implications of the spin-charge-$\eta$-spin separation for dynamical properties.
Thirdly, various concepts developed for one-dimensional systems in equilibrium can be extended to understand the physics of metastable states.
For example, extending the spin incoherent Luttinger liquids~\cite{Gregory2007RMP} may be helpful for understanding effectively cold, but not ultracold systems.

Finally but not least, our analytical and intuitive insights provide a useful reference for the study of photo-doped Mott insulators in higher dimensions, where the separation of spin, charge and $\eta$-spin is not expected, but a crossover from high-dimensional to one-dimensional behavior can occur in anisotropic systems.

\begin{acknowledgments}
We thank R. Arita for inspiring comments and A.~J. Millis, Z. Sun, D. Gole\v{z} and D. Baeriswyl for fruitful discussions.
This work was supported by Grant-in-Aid for Scientific Research from JSPS, KAKENHI Grant Nos. JP20K14412 (Y.M.), JP21H05017 (Y.M.), 21K03412 (S. T.), JP18K13509 (T.K.)
JP20H01849 (T.K.), JST CREST Grant No. JPMJCR1901 (Y.M.), JPMJCR19T3 (Y. M. and S. T.), and ERC Consolidator Grant No.~724103 (P.W.). 
\end{acknowledgments}

\bibliography{Ref_GGE}

\clearpage
\onecolumngrid
\section{Effective Hamiltonian}

In this study, we focus on metastable states in photo-doped Mott insulators with a large gap described by the one-band Hubbard model.
In such systems, the life-time of doublons and holons is known to be exponentially long for large on-site Coulomb interactions $U$, due to the lack of efficient recombination channels~\cite{Strohmaier2010PRL,Zala2013PRL,Mitrano2014,Sensarma2010PRB,Martin2011PRB,Zala2014PRB}.
To be more precise, since  the Hubbard model contains doublon-holon recombination terms, what is practically conserved are not the doublons and holons in the original Hamiltonian, but those dressed with virtual recombination processes.
The description in terms of such dressed doublons and holons can be obtained by the Schrieffer-Wolff (SW) transformation, which yields $\hat{H}_{\rm SW}$.
The SW transformation can be constructed for any order $n$, such that $\hat{H}_{\rm SW}$ has no recombination terms up to order $1/U^n$~\cite{MacDonald1988PRB}.
This implies that the life-time of the doublons and holons grows faster than any power of $U$ ~\cite{Rosch2008PRL}.
Because of this long life-time of the doublons and holons, the initial relaxation process (e.~g. after photo-doping) is the intraband relaxation due to scattering events and the dissipation of excess kinetic energy to the environment.
In other words, the system is cooled down and approaches a thermalized state of a given number of (dressed) doublons and holons.
To describe such a steady state it is natural to use the model obtained from the SW transformation which is truncated at a certain order, 
since the higher order terms are relevant only for the slow recombination processes~\cite{Takahashi2002PRB_A,Takahashi2002PRB,Takahashi2005PRB,Rosch2008PRL,Ishihara2011PRL,Ishihara2012PRB,Jiajun2020PRB,Jiajun2021PRB,Murakami2022}.

Note that our setup is analogous to that of photo-doped semiconductors, where conducting electrons and holes are generated. 
Here, the electron-hole recombination time is long, compared to the intraband relaxation, and the system first relaxes into a thermalized state with a given number of electrons and holes.
In the case of photo-doped semiconductors, the original Hamiltonian usually conserves the number of electrons and holes, and the steady states can be described as the equilibrium states of the model for a given number of electrons and holons at some effective temperature~\cite{haug_quantum_1990,Keldysh1986review,Asano2014JPSJ}.

To summarize, the basic assumptions of our setup are i) the long-life time of doublons and holons due to the large Mott gap and ii) the cooling of doublons and holons (intraband cooling) due to the coupling to the environment. The importance of ii) depends on the excitation protocol. Using appropriate excitation protocols, one can directly prepare states with little excess energy of the doublons and holons (effectively cold states)~\cite{Jiajun2020PRB}. Furthermore, in the context of cold atom systems, it has been discussed that essentially the same metastable state as the photo-doped Mott insulator can be prepared~\cite{Rosch2008PRL}.

The usage of the effective model obtained from the SW transformation for the description of photo-doped Mott insulators is supported by previous numerical studies. In a study based on the nonequilibrium dynamical mean-field theory (DMFT), the emergence of the $\eta$-paring phase in the photo-doped Hubbard model has been explicitly demonstrated and it was shown that its properties can be well explained by the effective model~\cite{Jiajun2020PRB}. It has also been reported that the photo-doped Hubbard model on the triangular lattice shows a chiral superconducting phase, which is also explained by the effective model~\cite{Jiajun2022arxiv}. Furthermore, the effective model was also derived for the two-band Hubbard model, and succeeded in explaining photo-induced phenomena experimentally observed in perovskite cobaltites~\cite{Ishihara2011PRL,Ishihara2012PRB}.

\section{Derivation of the wave function in the limit of $U\rightarrow \infty$ ($J_{\rm ex}\rightarrow0$)}

For the doped Hubbard model in equilibrium, Ogata and Shiba derived the exact form of the wave function in the large-$U$ limit. It can be expressed as $|\Psi \rangle = |\Psi_\text{charge}\rangle  |\Psi_\text{spin}\rangle$, which provides an intuitive picture of the spin-charge separation in this system. Originally, the Bethe ansatz solution was used to derive this expression~\cite{Ogata1990PRB}, while later on it was pointed out that the same result can be obtained using perturbation theory~\cite{Shiba1991}. In the following, we extend the latter strategy to derive the exact form of the wave function of the metastable photo-doped state in one-dimensional Mott insulators in the large-$U$ limit.

To be self-contained, we again write the effective Hamiltonian for the photo-doped system, 
\eqq{
 \hH_{\rm eff}  =   \hH_U + \hH_{\rm kin} + \hH_V +  \hH_{\rm spin,ex} +  \hH_{\rm dh,ex} + \hH_{U,\rm{shift}} + \hH_{\rm 3-site}. \label{eq:Heff}
}
Here, $\hH_U=U \sum_i (\hn_{i\uparrow}-\frac{1}{2}) ( \hn_{i\downarrow}-\frac{1}{2}) $, $ \hH_V \nonumber=V \sum_{\langle i,j\rangle } (\hn_i-1) (\hn_{j}-1)$
and $\hH_{\rm kin} = -t_{\rm hop}\sum_{\langle i,j\rangle,\sigma} \hat{\bar{n}}_{i,\bar{\sigma}} (\hc^\dagger_{i,\sigma}\hc_{j,\sigma} + h.c.)  \hat{\bar{n}}_{j,\bar{\sigma}} -t_{\rm hop}\sum_{\langle i,j\rangle,\sigma} \hn_{i,\bar{\sigma}} (\hc^\dagger_{i,\sigma}\hc_{j,\sigma} +h.c.)  \hn_{j,\bar{\sigma}}.$
The spin exchange term is 
$\hH_{\rm spin,ex}= J_{\rm ex} \sum_{\langle i,j\rangle} \hat{\bf  s}_i\cdot \hat{\bf  s}_{j}$ and 
the doublon-holon exchange term is $\hH_{\rm dh,ex}  = -J_{\rm ex}\sum_{\langle i,j\rangle} [ \heta^x_i \heta^x_{j} + \heta^y_i \heta^y_{j} + \heta^z_i \heta^z_{j}]$.
$\langle i,j\rangle$  denotes pairs of nearest-neighbor sites.
We use the same notation as in the main text.
The shift of the local interaction is described by $\hH_{U,{\rm shift}} =J_{\rm ex} \sum_i (\hn_{i\uparrow}-\tfrac{1}{2})(\hn_{i\downarrow}-\tfrac{1}{2})$.
  
\begin{figure}[t]
  \centering
    \hspace{-0.cm}
    \vspace{0.0cm}
\includegraphics[width=170mm]{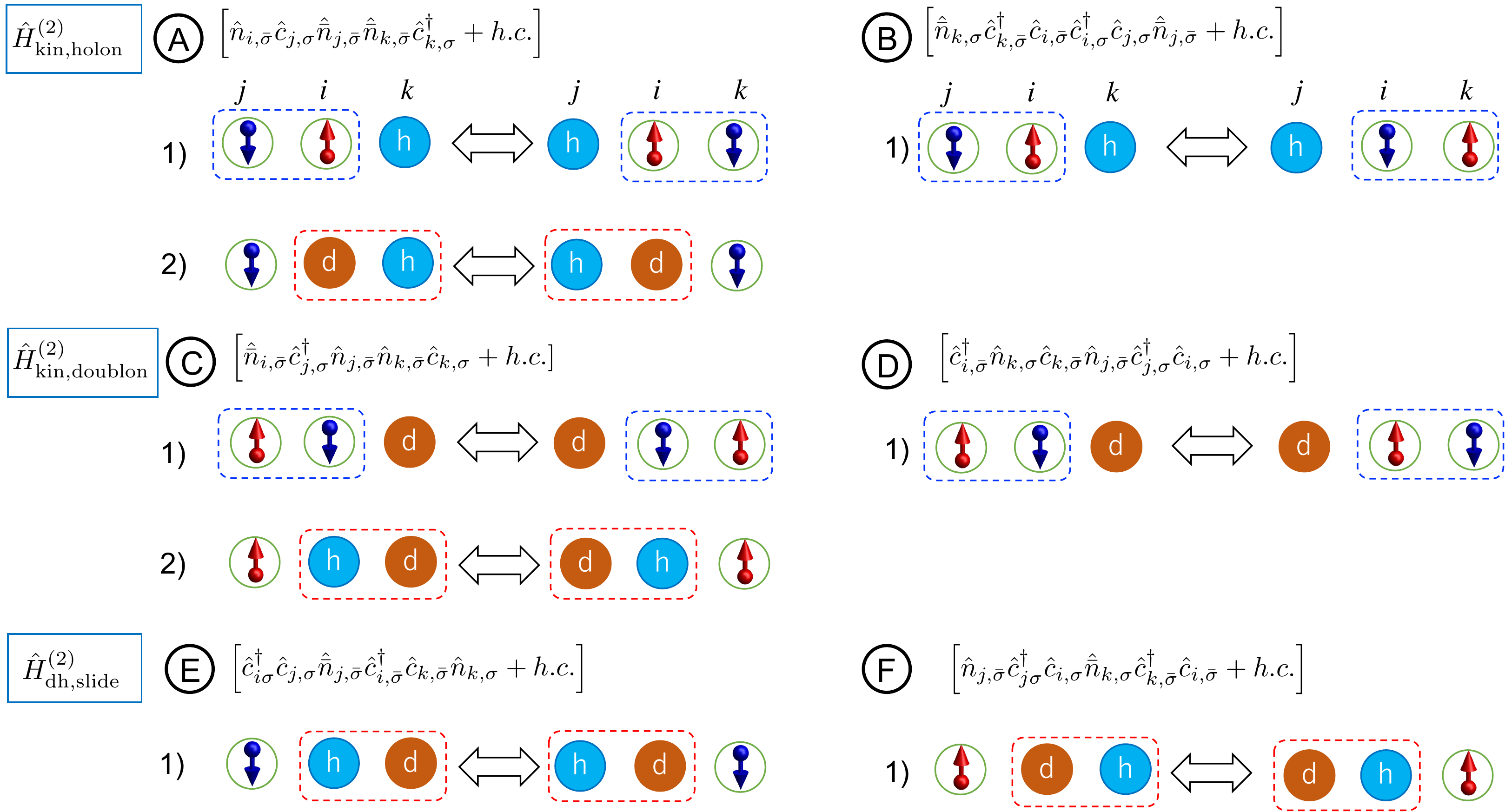} 
  \caption{Schematic picture of how each term in $\hH_{\rm 3-site}$ acts. We exemplify the cases that become nonzero for $\sigma=\,\downarrow$. For example, the first term in $\hH^{(2)}_{\rm kin,holon}$  ($ \hn_{i,\bar{\sigma}} \hc_{j,\sigma}\hat{\bar{n}}_{j,\bar{\sigma}} \hat{\bar{n}}_{k,\bar{\sigma}} \hc^\dagger_{k,\sigma} $) has two such cases 1) and 2), and application of this term induces the transition from the left states to the right states. The second term in $\hH^{(2)}_{\rm kin,holon}$, i.e. the hermitian conjugate of the first term, also has two cases 1) and 2), and its application induces  the transition from the right states to the left states (the opposite direction of the first term).
  The same holds for the other cases. Dashed rectangles indicate parts that are related to spin (blue) or $\eta$-spin (red) exchange couplings in the squeezed space. }
  \label{fig:H_3site}
\end{figure}

 The 3-site term can be expressed as $\hH_{\rm 3-site}\equiv \hH^{(2)}_{\rm kin,holon} +  \hH^{(2)}_{\rm kin,doub} +  \hH^{(2)}_{\rm dh,slide}$.
Here, $ \hH^{(2)}_{\rm kin,holon}$ and $\hH^{(2)}_{\rm kin,doub}$ describe correlated hoppings of holons and doublons, while 
$\hH^{(2)}_{\rm dh,slide}$ shifts the position of a doublon and a holon.
The explicit expressions are 
    \eqq{
  \hH^{(2)}_{\rm kin,holon}   =\frac{J_{\rm ex}}{4}\sum_{\langle k,i,j \rangle,\sigma} \Bigl[ 
    \hn_{i,\bar{\sigma}} \hc_{j,\sigma}\hat{\bar{n}}_{j,\bar{\sigma}} \hat{\bar{n}}_{k,\bar{\sigma}} \hc^\dagger_{k,\sigma} + 
    h.c.
    \Bigl]  - \frac{J_{\rm ex}}{4} \sum_{\langle k,i,j\rangle,\sigma}  \Bigl[\hat{\bar{n}}_{k,\sigma} \hc^\dagger_{k,\bar{\sigma}} \hc_{i,\bar{\sigma}} \hc^\dagger_{i,\sigma} \hc_{j,\sigma} \hat{\bar{n}}_{j,\bar{\sigma}}
    + h.c. \Bigl],
  }
 \eqq{
  \hH^{(2)}_{\rm kin,doublon}  
= \frac{J_{\rm ex}}{4}\sum_{\langle k,i,j\rangle,\sigma}  \Bigl[ \hat{\bar{n}}_{i,\bar{\sigma}} \hc^\dagger_{j,\sigma} \hn_{j,\bar{\sigma}} \hn_{k,\bar{\sigma}} \hc_{k,\sigma} 
+   h.c.]
 - \frac{J_{\rm ex}}{4} \sum_{\langle k,i,j\rangle,\sigma}  \Bigl[ \hc^\dagger_{i,\bar{\sigma}} \hn_{k,\sigma} \hc_{k,\bar{\sigma}} \hn_{j,\bar{\sigma}} \hc^\dagger_{j,\sigma} \hc_{i,\sigma}
  +  h.c.\Bigl],
 }
and 
 \eqq{
\hH^{(2)}_{\rm dh,slide}  
  = \frac{J_{\rm ex}}{4} \sum_{\langle k,i,j\rangle,\sigma} \Bigl[ \hc_{i\sigma}^\dagger \hc_{j,\sigma} \hat{\bar{n}}_{j,\bar{\sigma}} \hc^\dagger_{i,\bar{\sigma}}  \hc_{k,\bar{\sigma}}  \hat{n}_{k,\sigma}
   +   h.c. \Bigl] 
   + \frac{J_{\rm ex}}{4} \sum_{\langle k,i,j\rangle,\sigma}  \Bigl[\hn_{j,\bar{\sigma}} \hc_{j\sigma}^\dagger \hc_{i,\sigma}  \hat{\bar{n}}_{k,\sigma} \hc^\dagger_{k,\bar{\sigma}}  \hc_{i,\bar{\sigma}} 
  + h.c. \Bigl].
 }
 Here, $\langle k,i,j \rangle$ means that both  ($k$, $i$)  and  ($i$, $j$)  are pairs of neighboring sites. 
The summation is over all possible such combinations (without double counting, i.e. we identify $\langle k,i,j\rangle$ with $\langle j,i,k\rangle$).
We consider the open boundary condition.
In Fig.~\ref{fig:H_3site}, we schematically show how each term acts on appropriately chosen configurations.

We will consider the limit of $J_{\rm ex}\rightarrow 0$ with $V/J_{\rm ex}={\rm const}$. The system size is set to $L$, the number of singly occupied sites is $N_s$ and $N_{\eta}=L-N_s$. We also introduce the densities $n_s = N_s/L$ and $n_\eta = N_\eta/L$.
The system is assumed to be effectively cold and described by the ground state of the effective Hamiltonian.
As explained in the main text, to analyze the correlations between different degrees of freedom in this system, we introduce the Hilbert space
 \eqq{
 \mathcal{H}' = \Bigl\{ |\br\rangle|\bsig\rangle|\boldeta\rangle\equiv \Big(\prod_{r\in \br} \hc^\dagger_{r}\Big)|{\rm vac}\rangle|\bsig\rangle|\boldeta\rangle \text{: $\# \br =\# \bsig= N_s$ and $\# \boldeta = N_{\eta}$} \Bigl\},
 }
 and identify this with the original Hilbert space $\mathcal{H}$ using the unitary transformation $\hU:  \mathcal{H} \rightarrow  \mathcal{H}'$ given by
 \eqq{
 \hU \Bigl( \Bigl(\prod_{i=1}^{N_s} \hc^\dagger_{r_i,\sigma_i}\Bigl) \Bigl(\prod_{j=1}^{N_{\eta}} \ha^\dagger_{\bar{r}_j,\eta_j}\Bigl)|{\rm vac}\rangle  \Bigl) = \Big(\prod_{r\in \br} \hc^\dagger_{r}\Big)|{\rm vac}\rangle|\bsig\rangle|\boldeta\rangle.
 }
Here $\br = \{r_{N_s}, \cdots r_1\}$ with $L\geq r_{N_s}>r_{N_s-1}>\cdots > r_{1}\geq 1$, $\bar{\br} = \{\bar{r}_{N_{\eta}}, \cdots \bar{r}_1\}$ with $L\geq \bar{r}_{N_{\eta}}>\bar{r}_{N_{\eta}-1}>\cdots > \bar{r}_{1}\geq 1$, and $\br \cup \bar{\br} = \{L,L-1,\cdots,1\}$. $\hc^\dagger_r$ is the creation operator of a spinless fermion (SF), $\eta =\, \uparrow,\downarrow$ labels doublons and holons, $\ha^\dagger_{\bar{r},\uparrow} = (-)^{\bar{r}} \hc^\dagger_{\bar{r}\downarrow} \hc^\dagger_{\bar{r}\uparrow}$ and $\ha^\dagger_{\bar{r},\downarrow} = 1$.
 
The relevant statements are the following:
 \begin{enumerate}[(i)]
 \item $ \hU \hH_{\rm kin}  \hU^\dagger  = -t_{\rm hop}\sum_{\langle i,j\rangle} (\hc^\dagger_{i}\hc_{j} + h.c.)$ $(\equiv \hH_{\rm 0,SF})$.
 This means that, in the representation of $\mathcal{H}'$, the wave function of the 0th order model (without $\mathcal{O}(J_{\rm ex})$ terms) can be expressed as $|\Psi^{\rm GS}_{\rm SF}\rangle|\Psi_{\sigma,\eta}\rangle$, where $|\Psi^{\rm GS}_{\rm SF}\rangle$ 
 is the ground state of $ \hH_{\rm 0,SF}$ and $|\Psi_{\sigma,\eta}\rangle$ is an arbitrary spin and $\eta$-spin wave function.
 
 \item When the $\mathcal{O}(J_{\rm ex})$ terms ($\equiv \hH_{\rm perturb}$) are projected to $|\Psi^{\rm GS}_{\rm SF}\rangle|\bsig\rangle|\boldeta\rangle$, the Hamiltonian is separated into a spin Hamiltonian and an $\eta$-spin Hamiltonian. Thus, within lowest-order degenerate perturbation theory, the wave function in $\mathcal{H}'$ can be expressed as $|\Psi^{\rm GS}_{\rm SF}\rangle|\Psi^{\rm GS}_{\sigma}\rangle|\Psi^{\rm GS}_{\eta}\rangle$.
$ |\Psi^{\rm GS}_{\sigma}\rangle$ is the ground state of the Heisenberg model in the squeezed spin space, and $|\Psi^{\rm GS}_{\eta}\rangle$ is the ground state of the XXZ model in the squeezed $\eta$-spin space.
 \end{enumerate}
 
 In the following subsections, we provide detailed calculations to support these statements. 
  
 \subsection{Point (i): Proof of $ \hU \hH_{\rm kin}  \hU^\dagger  = -t_{\rm hop}\sum_{\langle i,j\rangle} (\hc^\dagger_{i}\hc_{j} + h.c.)$}
 In order to prove $ \hU \hH_{\rm kin}  \hU^\dagger  = -t_{\rm hop}\sum_{\langle i,j\rangle} (\hc^\dagger_{i}\hc_{j} + h.c.)$, we focus on $\hV_r \equiv \sum_{\sigma} \hat{\bar{n}}_{r,\bar{\sigma}} (\hc^\dagger_{r,\sigma}\hc_{r+1,\sigma} + h.c.) \hat{\bar{n}}_{r+1,\bar{\sigma}} + \sum_{\sigma} \hn_{r,\bar{\sigma}} (\hc^\dagger_{r,\sigma}\hc_{r+1,\sigma} + h.c.) \hn_{r+1,\bar{\sigma}}$ and the expected corresponding term in $\mathcal{H}'$, i.e. $\hV_{r, {\rm SF}} \equiv (\hc^\dagger_{r}\hc_{r+1} + h.c.)$.
 We have the following properties.
 \begin{enumerate}
\item $r\in \br \land r+1\in  \br$: \\
In this case, $\hV_r U^\dagger |\br\rangle|\bsig\rangle|\boldeta\rangle= 0$ and $ \hat{U}\hV_r U^\dagger |\br\rangle|\bsig\rangle|\boldeta\rangle = \hV_{r, {\rm SF}} |\br\rangle|\bsig\rangle|\boldeta\rangle$.
\item $r\notin \br \land r+1\notin \br $: \\
In this case, $\hV_r U^\dagger |\br\rangle|\bsig\rangle|\boldeta\rangle= 0$ and $ \hat{U}\hV_r U^\dagger |\br\rangle|\bsig\rangle|\boldeta\rangle = \hV_{r, {\rm SF}} |\br\rangle|\bsig\rangle|\boldeta\rangle$.
\item $r\in \br \land r+1\notin \br$: 
\begin{enumerate}[(a)]
\item If $\ha^\dagger_{r+1,\eta} = \ha^\dagger_{r+1,\downarrow}$,
\eqq{
\hV_r U^\dagger |\br\rangle|\bsig\rangle|\boldeta\rangle  &=  \sum_{\sigma} \hat{\bar{n}}_{r,\bar{\sigma}} \hc^\dagger_{r+1,\sigma} \hc_{r,\sigma} \hat{\bar{n}}_{r+1,\bar{\sigma}} (\cdots \ha^\dagger_{r+1,\downarrow}\hc^\dagger_{r\sigma'}\cdots) |{\rm vac}\rangle \nonumber \\
&= (\cdots   \hc^\dagger_{r+1\sigma'} \ha^\dagger_{r,\downarrow}\cdots) |{\rm vac}\rangle  = \hat{U}^\dagger \hV_{r, {\rm SF}} |\br\rangle|\bsig\rangle|\boldeta\rangle  .
}
\item If $\ha^\dagger_{r+1,\eta} = \ha^\dagger_{r+1,\uparrow}$,
\eqq{
\hV_r U^\dagger |\br\rangle|\bsig\rangle|\boldeta\rangle  &=  \sum_{\sigma} \hat{n}_{r,\bar{\sigma}} \hc^\dagger_{r,\sigma} \hc_{r+1,\sigma} \hn_{r+1,\bar{\sigma}} (\cdots (-)^{r+1} \hc^\dagger_{r+1\downarrow} \hc^\dagger_{r+1\uparrow}\hc^\dagger_{r\sigma'}\cdots) |{\rm vac}\rangle \nonumber \\
&= (\cdots   \hc^\dagger_{r+1\sigma'} \ha^\dagger_{r,\uparrow}\cdots) |{\rm vac}\rangle  = \hat{U}^\dagger \hV_{r, {\rm SF}} |\br\rangle|\bsig\rangle|\boldeta\rangle  .
}
\end{enumerate}
\item $r\notin \br \land r+1\in \br$:
\begin{enumerate}[(a)]
\item If $\ha^\dagger_{r,\eta} = \ha^\dagger_{r,\downarrow}$,
\eqq{
\hV_r U^\dagger |\br\rangle|\bsig\rangle|\boldeta\rangle  &=  \sum_{\sigma} \hat{\bar{n}}_{r,\bar{\sigma}} \hc^\dagger_{r,\sigma} \hc_{r+1,\sigma} \hat{\bar{n}}_{r+1,\bar{\sigma}} (\cdots \hc^\dagger_{r+1\sigma'} \ha^\dagger_{r,\downarrow}\cdots) |{\rm vac}\rangle \nonumber \\
&= (\cdots \ha^\dagger_{r+1,\downarrow}  \hc^\dagger_{r\sigma'} \cdots) |{\rm vac}\rangle  = \hat{U}^\dagger \hV_{r, {\rm SF}} |\br\rangle|\bsig\rangle|\boldeta\rangle  .
}
\item If $\ha^\dagger_{r,\eta} = \ha^\dagger_{r,\uparrow}$,
\eqq{
\hV_r U^\dagger |\br\rangle|\bsig\rangle|\boldeta\rangle  &=  \sum_{\sigma} \hn_{r,\bar{\sigma}} \hc^\dagger_{r+1,\sigma} \hc_{r,\sigma} \hn_{r+1,\bar{\sigma}} (\cdots \hc^\dagger_{r+1\sigma'} (-)^{r} \hc^\dagger_{r\downarrow} \hc^\dagger_{r\uparrow}\cdots) |{\rm vac}\rangle \nonumber \\
&= (\cdots   \ha^\dagger_{r+1,\uparrow} \hc^\dagger_{r\sigma'} \cdots) |{\rm vac}\rangle  = \hat{U}^\dagger \hV_{r,{\rm SF}} |\br\rangle|\bsig\rangle|\boldeta\rangle .
}
\end{enumerate}
\end{enumerate}
Thus, we obtain $ \hU \hV_r  \hU^\dagger  = \hV_{r, {\rm SF}}$ and $ \hU \hH_{\rm kin}  \hU^\dagger  = -t_{\rm hop}\sum_{\langle i,j\rangle} (\hc^\dagger_{i} \hc_{j} + h.c.)$.

\subsection{Point (ii) : Expression for  $ \langle\boldeta'|\langle \bsig' |\langle \Psi^{\rm GS}_{\rm SF} | \hU \hH_{\rm perturb} \hU^\dagger|\Psi^{\rm GS}_{\rm SF}\rangle|\bsig\rangle|\boldeta\rangle$}
Here we want to evaluate $ \langle\boldeta'|\langle \bsig' |\langle \Psi^{\rm GS}_{\rm SF} | \hU \hH_{\rm perturb} \hU^\dagger|\Psi^{\rm GS}_{\rm SF}\rangle|\bsig\rangle|\boldeta\rangle$.
First, we focus on the two site terms $\hH_{2-{\rm site}}\equiv \hH_{\rm spin,ex}+\hH_{\rm dh,ex}+\hH_V$. To this end, we introduce $\hV^\sigma_{Z}=\sum_r \hs^z_{r+1} \hs^z_r$, $\hV^\sigma_{XY}=\sum_r [\hs^+_{r+1} \hs^-_r + \hs^-_{r+1} \hs^+_r ]$, and $\hV^\eta_{Z}=\sum_r \heta^z_{r+1} \heta^z_r$, $\hV^\eta_{XY}=\sum_r [\heta^+_{r+1} \heta^-_r + \heta^-_{r+1} \heta^+_r ]$ in the original Hilbert space $\mathcal{H}$ and  consider the corresponding expression in $\mathcal{H'}$.
In the following, we introduce $\psi_{\rm c}({\bf r})$ via $|\Psi^{\rm GS}_{\rm SF}\rangle = \sum_{{\bf r}\in \Lambda} \psi_{\rm c}({\bf r})|{\bf r}\rangle$.
Note that ${\bf r}$ is a set of $N_s$ position indices and $\Lambda$ is the group of such sets.

\begin{enumerate}
\item $\hV^\sigma_{Z}=\sum_r \hs^z_{r+1} \hs^z_r$: 

Firstly, we can show that
\eqq{
& \hU \hV^\sigma_{Z} \hU^\dagger  |\br \rangle|\bsig\rangle|\boldeta\rangle = \hU \hV^\sigma_{Z} \Bigl(\prod_{i=1}^{N_s} \hc^\dagger_{r_i,\sigma_i}\Bigl) \Bigl(\prod_{j=1}^{N_{\eta}} \ha^\dagger_{\bar{r}_j,\eta_j}\Bigl)|{\rm vac}\rangle   \nonumber \\
& =\hU \sum_i \delta_{r_{i +1},r_i+1} \frac{\sigma_{i+1}\sigma_i}{4}(\cdots \hc^\dagger_{r_{i+1},\sigma_{i+1}} \hc^\dagger_{r_i,\sigma_i}  \cdots) |{\rm vac}\rangle  
= \sum_i \hn_{r_i} \hn_{r_i +1} [\hs^z_{i+1} \hs^z_i ]  |\br \rangle|\bsig\rangle|\boldeta\rangle.
}
Here, $\sigma_i$ in the numerator takes the value 1 ($-1$) if the index $\sigma_i$ is $\uparrow$ ($\downarrow$).
Then we have 
\eqq{
&\hU \hV^\sigma_{Z} \hU^\dagger  |\Psi^{\rm GS}_{\rm SF}\rangle |\bsig\rangle|\boldeta\rangle = \sum_{\br \in \Lambda} \psi_{\rm c}(\br)  \sum_i \hn_{r_i} \hn_{r_i +1} \hs^z_{i+1} \hs^z_i   |\br \rangle|\bsig\rangle|\boldeta\rangle \nonumber \\
& = \sum_i  \Bigl(\sum_{\br \in \Lambda} \psi_{\rm c}(\br)  \hn_{r_i} \hn_{r_i +1}  |\br \rangle\Bigl) \hs^z_{i+1} \hs^z_i  |\bsig\rangle|\boldeta\rangle.
}
Therefore, we obtain
\eqq{
\langle\boldeta'|\langle \bsig' |\langle \Psi^{\rm GS}_{\rm SF} | \hU \hV^\sigma_{Z} \hU^\dagger|\Psi^{\rm GS}_{\rm SF}\rangle|\bsig\rangle|\boldeta\rangle = \delta_{\boldeta',\boldeta}
\sum_i \Bigl( \langle \Psi^{\rm GS}_{\rm SF}| \sum_{\br \in \Lambda} \psi_{\rm c}(\br)  \hn_{r_i} \hn_{r_i +1}  |\br \rangle  \Bigl)  \langle \bsig'|\hs^z_{i+1} \hs^z_i |\bsig\rangle,
}
which shows that this operator only acts on the spin sector.
When the system is large enough, the boundary effect can be neglected and we can replace $\langle \Psi^{\rm GS}_{\rm SF}| \sum_{\br \in \Lambda} \psi_{\rm c}(\br)  \hn_{r_i} \hn_{r_i +1}  |\br \rangle$
by $\tilde{x} \equiv \frac{1}{N_s} \sum_r  \langle \Psi^{\rm GS}_{\rm SF}| \hn_{r} \hn_{r +1}|\Psi^{\rm GS}_{\rm SF}\rangle$~\cite{Shiba1991}.
\item $\hV^\sigma_{XY}=\sum_r [\hs^+_{r+1} \hs^-_r + \hs^-_{r+1} \hs^+_r ]$:

Firstly, we can show
\eqq{
& \hU \hV^\sigma_{XY} \hU^\dagger  |\br \rangle|\bsig\rangle|\boldeta\rangle = \hU \hV^\sigma_{XY} \Bigl(\prod_{i=1}^{N_s} \hc^\dagger_{r_i,\sigma_i}\Bigl) \Bigl(\prod_{j=1}^{N_{\eta}} \ha^\dagger_{\bar{r}_j,\eta_j}\Bigl)|{\rm vac}\rangle   \nonumber \\
& =\hU \sum_i \delta_{r_{i +1},r_i+1} \delta_{\sigma_{i+1},\bar{\sigma}_i}(\cdots \hc^\dagger_{r_{i+1},\bar{\sigma}_{i+1}} \hc^\dagger_{r_i,\bar{\sigma}_i}  \cdots) |{\rm vac}\rangle  
= \sum_i \hn_{r_i} \hn_{r_i +1} [\hs^+_{i+1} \hs^-_i + \hs^-_{i+1} \hs^+_i ]  |\br \rangle|\bsig\rangle|\boldeta\rangle.
}
Here $\bar{\sigma}$ denotes the opposite of $\sigma$.
Then we have 
\eqq{
&\hU \hV^\sigma_{XY} \hU^\dagger  |\Psi^{\rm GS}_{\rm SF}\rangle |\bsig\rangle|\boldeta\rangle = \sum_{\br \in \Lambda} \psi_{\rm c}(\br)  \sum_i \hn_{r_i} \hn_{r_i +1} [\hs^+_{i+1} \hs^-_i + \hs^-_{i+1} \hs^+_i ]  |\br \rangle|\bsig\rangle|\boldeta\rangle \nonumber \\
& = \sum_i  \Big( \sum_{\br \in \Lambda} \psi_{\rm c}(\br)  \hn_{r_i} \hn_{r_i +1}  \Big) ([\hs^+_{i+1} \hs^-_i + \hs^-_{i+1} \hs^+_i ] |\bsig\rangle)|\boldeta\rangle.
}
Therefore, we obtain 
\eqq{
\langle\boldeta'|\langle \bsig' |\langle \Psi^{\rm GS}_{\rm SF} | \hU \hV^\sigma_{XY} \hU^\dagger|\Psi^{\rm GS}_{\rm SF}\rangle|\bsig\rangle|\boldeta\rangle = \delta_{\boldeta',\boldeta}
\sum_i \Bigl( \langle  \Psi^{\rm GS}_{\rm SF}| \sum_{\br \in \Lambda} \psi_{\rm c}(\br)  \hn_{r_i} \hn_{r_i +1}  |\br \rangle  \Bigl)  \langle \bsig'|[\hs^+_{i+1} \hs^-_i + \hs^-_{i+1} \hs^+_i ] |\bsig\rangle,
}
which shows that this part only applies to the spin sector.
As for $\hV^\sigma_{Z}$, considering a large size system, we can replace $( \langle  \Psi^{\rm GS}_{\rm SF}| \sum_{\br \in \Lambda} \psi_{\rm c}(\br)  \hn_{r_i} \hn_{r_i +1}  |\br \rangle  ) $
with $\tilde{x} = \frac{1}{N_s} \sum_r  \langle \Psi^{\rm GS}_{\rm SF}| \hn_{r} \hn_{r +1}|\Psi^{\rm GS}_{\rm SF}\rangle$.
\item $\hV^\eta_{Z}=\sum_r \heta^z_{r+1} \heta^z_r$:

Firstly, we can show
\eqq{
& \hU \hV^\eta_{Z}\hU^\dagger  |\br \rangle|\bsig\rangle|\boldeta\rangle = \hU \hV^\eta_{Z} \Bigl(\prod_{i=1}^{N_s} \hc^\dagger_{r_i,\sigma_i}\Bigl) \Bigl(\prod_{j=1}^{N_{\eta}} \ha^\dagger_{\bar{r}_j,\eta_j}\Bigl)|{\rm vac}\rangle   \nonumber \\
& =\hU \sum_j \delta_{\bar{r}_{j +1},\bar{r}_j+1} \frac{\eta_{j+1} \eta_{j}}{4} (\cdots \ha^\dagger_{\bar{r}_{j+1},\eta_{j+1}} \ha^\dagger_{\bar{r}_{j},\eta_{j}}  \cdots) |{\rm vac}\rangle  
= \sum_j \hat{\bar{n}}_{\bar{r}_j} \hat{\bar{n}}_{\bar{r}_j +1} \heta^z_{j+1} \heta^z_j |\br \rangle|\bsig\rangle|\boldeta\rangle.
}
Here, $\eta_i$ in the numerator takes the value 1 ($-1$) if the index $\eta_i$ is $\uparrow$ ($\downarrow$).
Then we have 
\eqq{
&\hU \hV^\eta_{Z} \hU^\dagger  |\Psi^{\rm GS}_{\rm SF}\rangle |\bsig\rangle|\boldeta\rangle = \sum_{\br \in \Lambda} \psi_{\rm c}(\br)  \sum_j \hat{\bar{n}}_{\bar{r}_j} \hat{\bar{n}}_{\bar{r}_j +1} \heta^z_{j+1} \heta^z_j|\br \rangle|\bsig\rangle|\boldeta\rangle \nonumber \\
& = \sum_j (\sum_{\br \in \Lambda} \psi_{\rm c}(\br)  \hat{\bar{n}}_{\bar{r}_j} \hat{\bar{n}}_{\bar{r}_j +1}  |\br \rangle)  |\bsig\rangle (\heta^z_{j+1} \heta^z_j|\boldeta\rangle).
}
Therefore we obtain
\eqq{
\langle\boldeta'|\langle \bsig' |\langle \Psi^{\rm GS}_{\rm SF} | \hU \hV^\eta_{Z} \hU^\dagger|\Psi^{\rm GS}_{\rm SF}\rangle|\bsig\rangle|\boldeta\rangle = \delta_{\bsig',\bsig}
\sum_j \Bigl( \langle \Psi^{\rm GS}_{\rm SF}| \sum_{\br \in \Lambda} \psi_{\rm c}(\br)  \hat{\bar{n}}_{\bar{r}_j} \hat{\bar{n}}_{\bar{r}_j +1}  |\br \rangle  \Bigl)  \langle \boldeta'| \heta^z_{j+1} \heta^z_j |\boldeta\rangle,
}
which shows that this operator only acts on the $\eta$-spin sector. For a large enough system, we have $ \langle \Psi^{\rm GS}_{\rm SF}| \sum_{\br \in \Lambda} \psi_{\rm c}(\br)  \hat{\bar{n}}_{\bar{r}_j} \hat{\bar{n}}_{\bar{r}_j +1}  |\br \rangle \simeq
 \frac{1}{N_{\eta}} \sum_j  \langle \Psi^{\rm GS}_{\rm SF}| \sum_{\br \in \Lambda} \psi_{\rm c}(\br)  \hat{\bar{n}}_{\bar{r}_j} \hat{\bar{n}}_{\bar{r}_j +1}  |\br \rangle=  \frac{1}{N_{\eta}} \sum_r  \langle \Psi^{\rm GS}_{\rm SF}| \sum_{\br \in \Lambda} \psi_{\rm c}(\br)  \hat{\bar{n}}_{r} \hat{\bar{n}}_{r +1}  |\br \rangle$ $(\equiv \tilde{y})$.
 
 \clearpage
\item $\hV^\eta_{XY}=\sum_r [\heta^+_{r+1} \heta^-_r + \heta^-_{r+1} \heta^+_r ]$:

Firstly, we can show
\eqq{
& \hU \hV^\eta_{XY}\hU^\dagger  |\br \rangle|\bsig\rangle|\boldeta\rangle = \hU \hV^\eta_{XY} \Bigl(\prod_{i=1}^{N_s} \hc^\dagger_{r_i,\sigma_i}\Bigl) \Bigl(\prod_{j=1}^{N_{\eta}} \ha^\dagger_{\bar{r}_j,\eta_j}\Bigl)|{\rm vac}\rangle   \nonumber \\
& =\hU \sum_j \delta_{\bar{r}_{j +1},\bar{r}_j+1} \delta_{\eta_{j+1},\bar{\eta}_j}(\cdots \ha^\dagger_{\bar{r}_{j+1},\bar{\eta}_{j+1}} \ha^\dagger_{\bar{r}_{j},\bar{\eta}_{j}}  \cdots) |{\rm vac}\rangle  
= \sum_j \hat{\bar{n}}_{\bar{r}_j} \hat{\bar{n}}_{\bar{r}_j +1} [\heta^+_{j+1} \heta^-_j + \heta^-_{j+1} \heta^+_j ] |\br \rangle|\bsig\rangle|\boldeta\rangle.
}
Here $\bar{\eta}$ denotes the opposite of $\eta$.
Then we have 
\eqq{
&\hU \hV^\eta_{XY} \hU^\dagger  |\Psi^{\rm GS}_{\rm SF}\rangle |\bsig\rangle|\boldeta\rangle = \sum_{\br \in \Lambda} \psi_{\rm c}(\br)  \sum_j \hat{\bar{n}}_{\bar{r}_j} \hat{\bar{n}}_{\bar{r}_j +1} [\heta^+_{j+1} \heta^-_j + \heta^-_{j+1} \heta^+_j ]|\br \rangle|\bsig\rangle|\boldeta\rangle \nonumber \\
& = \sum_j (\sum_{\br \in \Lambda} \psi_{\rm c}(\br)  \hat{\bar{n}}_{\bar{r}_j} \hat{\bar{n}}_{\bar{r}_j +1}  |\br \rangle)  |\bsig\rangle ([\heta^+_{j+1} \heta^-_j + \heta^-_{j+1} \heta^+_j ]|\boldeta\rangle).
}
Therefore, we obtain 
\eqq{
&\langle\boldeta'|\langle \bsig' |\langle \Psi^{\rm GS}_{\rm SF} | \hU \hV^\eta_{XY} \hU^\dagger|\Psi^{\rm GS}_{\rm SF}\rangle|\bsig\rangle|\boldeta\rangle = \delta_{\bsig',\bsig}
\sum_j \Bigl( \langle \Psi^{\rm GS}_{\rm SF}| \sum_{\br \in \Lambda} \psi_{\rm c}(\br)  \hat{\bar{n}}_{\bar{r}_j} \hat{\bar{n}}_{\bar{r}_j +1}  |\br \rangle  \Bigl)  \langle \boldeta'|  [\heta^+_{j+1} \heta^-_j + \heta^-_{j+1} \heta^+_j ] |\boldeta\rangle,
}
which shows that this operator only acts on the $\eta$-spin sector. As in the case of $\hV^\eta_{Z}$, $ \langle \Psi^{\rm GS}_{\rm SF}| \sum_{\br \in \Lambda} \psi_{\rm c}(\br)  \hat{\bar{n}}_{\bar{r}_j} \hat{\bar{n}}_{\bar{r}_j +1}  |\br \rangle$ can be replaced by $\tilde{y}$ for large systems.
\end{enumerate}

Next we consider the 3-site terms. To this end, we separate $\hH_{\rm 3-site}$ into $\hH^{(2)}_{\rm kin,holon} +  \hH^{(2)}_{\rm kin,doub}$ and  $ \hH^{(2)}_{\rm dh,slide}$ and evaluate the projections to $|\Psi^{\rm GS}_{\rm SF}\rangle|\bsig\rangle|\boldeta\rangle$ separately, see also Fig.~\ref{fig:H_3site}.

$\hH^{(2)}_{\rm kin,holon} +  \hH^{(2)}_{\rm kin,doub}$ yields the following four cases.
\begin{enumerate}
\item $\hV_1\equiv \sum_{r,\sigma} \hn_{r,\bar{\sigma}} \hc_{r+1,\sigma} \hat{\bar{n}}_{r+1,\bar{\sigma}}  \hat{\bar{n}}_{r-1,\bar{\sigma}} \hc_{r-1,\sigma}^\dagger 
+\sum_{r,\sigma} \hat{\bar{n}}_{r,\bar{\sigma}} \hc^\dagger_{r+1,\sigma} \hn_{r+1,\bar{\sigma}}  \hn_{r-1,\bar{\sigma}} \hc_{r-1,\sigma}$:

 In this case, we have 
\eqq{
\langle\boldeta'|\langle \bsig' |\langle \Psi^{\rm GS}_{\rm SF} | \hU \hV_{1} \hU^\dagger|\Psi^{\rm GS}_{\rm SF}\rangle|\bsig\rangle|\boldeta\rangle =
 -\delta_{\boldeta',\boldeta} \tilde{x}'\sum_i \langle \bsig' | \hs^+_{i+1} \hs^-_{i}+ \hs^-_{i+1} \hs^+_{i} | \bsig \rangle 
 + \delta_{\bsig',\bsig} \tilde{y}'\sum_j \langle \boldeta' | \heta^+_{j} \heta^-_{j-1}+\heta^-_{j} \heta^+_{j-1} | \boldeta \rangle. \label{eq:V1}
}
\item $\hV_2\equiv \sum_{r,\sigma} \hn_{r,\bar{\sigma}} \hc_{r-1,\sigma} \hat{\bar{n}}_{r-1,\bar{\sigma}}  \hat{\bar{n}}_{r+1,\bar{\sigma}} \hc_{r+1,\sigma}^\dagger 
+\sum_{r,\sigma} \hat{\bar{n}}_{r,\bar{\sigma}} \hc^\dagger_{r-1,\sigma} \hn_{r-1,\bar{\sigma}}  \hn_{r+1,\bar{\sigma}} \hc_{r+1,\sigma}$:

 In this case, we have 
\eqq{
\langle\boldeta'|\langle \bsig' |\langle \Psi^{\rm GS}_{\rm SF} | \hU \hV_{2} \hU^\dagger|\Psi^{\rm GS}_{\rm SF}\rangle|\bsig\rangle|\boldeta\rangle =
 -\delta_{\boldeta',\boldeta} \tilde{x}'\sum_i \langle \bsig' | \hs^+_i \hs^-_{i-1}+ \hs^-_i \hs^+_{i-1} | \bsig \rangle 
 + \delta_{\bsig',\bsig} \tilde{y}'\sum_j \langle \boldeta' | \heta^+_{j+1} \heta^-_{j}+\heta^-_{j+1} \heta^+_{j} | \boldeta \rangle.
}
\item $\hV_3\equiv \sum_{r,\sigma} \hat{\bar{n}}_{r+1,\sigma} \hc^\dagger_{r+1,\bar{\sigma}} \hc^\dagger_{r,\sigma} \hc_{r,\bar{\sigma}}\hc_{r-1,\sigma}\hat{\bar{n}}_{r-1,\bar{\sigma}}
 + \sum_{r,\sigma} \hn_{r-1,\bar{\sigma}} \hc^\dagger_{r-1,\sigma} \hc^\dagger_{r,\bar{\sigma}} \hc_{r,\sigma}  \hc_{r+1,\bar{\sigma}} \hn_{r+1,\sigma}$:
 
  In this case, we have 
\eqq{
\langle\boldeta'|\langle \bsig' |\langle \Psi^{\rm GS}_{\rm SF} | \hU \hV_{3} \hU^\dagger|\Psi^{\rm GS}_{\rm SF}\rangle|\bsig\rangle|\boldeta\rangle = -\delta_{\boldeta',\boldeta}
\tilde{x}'\sum_i \langle \bsig' | 2(\hs^z_i \hs^z_{i+1}-\frac{1}{4})| \bsig \rangle.
}
\item $\hV_4\equiv \sum_{r,\sigma} \hat{\bar{n}}_{r-1,\sigma} \hc^\dagger_{r-1,\bar{\sigma}} \hc^\dagger_{r,\sigma} \hc_{r,\bar{\sigma}}\hc_{r+1,\sigma}\hat{\bar{n}}_{r+1,\bar{\sigma}}
 + \sum_{r,\sigma} \hn_{r+1,\bar{\sigma}} \hc^\dagger_{r+1,\sigma} \hc^\dagger_{r,\bar{\sigma}} \hc_{r,\sigma}  \hc_{r-1,\bar{\sigma}} \hn_{r-1,\sigma}$:
 
 In this case, we have 
\eqq{
\langle\boldeta'|\langle \bsig' |\langle \Psi^{\rm GS}_{\rm SF} | \hU \hV_{4} \hU^\dagger|\Psi^{\rm GS}_{\rm SF}\rangle|\bsig\rangle|\boldeta\rangle = -\delta_{\boldeta',\boldeta}
\tilde{x}'\sum_i \langle \bsig' | 2(\hs^z_i \hs^z_{i+1}-\frac{1}{4})| \bsig \rangle.
}
\end{enumerate}

$ \hH^{(2)}_{\rm dh,slide}$ yields the following two cases.
\begin{enumerate}
\item $\hV_5\equiv \sum_{r,\sigma}  \hat{\bar{n}}_{r+1,\bar{\sigma}} \hc_{r+1,\sigma}\hc^\dagger_{r,\bar{\sigma}} \hc^\dagger_{r,\sigma} \hc_{r-1,\bar{\sigma}}\hn_{r-1,\sigma}
+\sum_{r,\sigma} \hn_{r+1,\bar{\sigma}} \hc^\dagger_{r+1,\sigma} \hc_{r,\bar{\sigma}} \hc_{r,\sigma} \hc_{r-1,\bar{\sigma}}^\dagger \hat{\bar{n}}_{r-1,\sigma}$:

In this case, we have 
\eqq{
\langle\boldeta'|\langle \bsig' |\langle \Psi^{\rm GS}_{\rm SF} | \hU \hV_{5} \hU^\dagger|\Psi^{\rm GS}_{\rm SF}\rangle|\bsig\rangle|\boldeta\rangle = \delta_{\bsig',\bsig}
\tilde{y}'\sum_j \langle \boldeta' | 2(\heta^z_j \heta^z_{j+1}-\frac{1}{4})| \boldeta \rangle.
}
\item $\hV_6\equiv \sum_{r,\sigma}  \hat{\bar{n}}_{r-1,\bar{\sigma}} \hc_{r-1,\sigma}\hc^\dagger_{r,\bar{\sigma}} \hc^\dagger_{r,\sigma} \hc_{r+1,\bar{\sigma}} \hn_{r+1,\sigma}
+\sum_{r,\sigma} \hn_{r-1,\bar{\sigma}} \hc^\dagger_{r-1,\sigma} \hc_{r,\bar{\sigma}} \hc_{r,\sigma} \hc_{r+1,\bar{\sigma}}^\dagger \hat{\bar{n}}_{r+1,\sigma}$:

In this case, we have 
\eqq{
\langle\boldeta'|\langle \bsig' |\langle \Psi^{\rm GS}_{\rm SF} | \hU \hV_{6} \hU^\dagger|\Psi^{\rm GS}_{\rm SF}\rangle|\bsig\rangle|\boldeta\rangle = \delta_{\bsig',\bsig}
\tilde{y}'\sum_j \langle \boldeta' | 2(\heta^z_j \heta^z_{j+1}-\frac{1}{4})| \boldeta \rangle.
}
\end{enumerate}
Here, $\tilde{x}' = \frac{1}{N_s}\sum_r  \langle \Psi^{\rm GS}_{\rm SF}| \hc^\dagger_{r-1} \hn_r \hc_{r+1}|\Psi^{\rm GS}_{\rm SF}\rangle$ and
  $\tilde{y}' = -\frac{1}{N_{\eta}}\sum_r  \langle \Psi^{\rm GS}_{\rm SF}| \hc^\dagger_{r-1} \hat{\bar{n}}_r \hc_{r+1}|\Psi^{\rm GS}_{\rm SF}\rangle$.
 The derivations of these expressions are straightforward but lengthy. Thus, we just show the proof for the case of $\hV_1$. The other cases can be treated in a similar manner.

Firstly we want to know the expression for $\hU\hV_1\hU^\dagger|{\bf r}\rangle|\bsig\rangle|\boldeta\rangle$. 
To this end, we introduce $\hV_1(r)\equiv \sum_{\sigma} \hn_{r,\bar{\sigma}} \hc_{r+1,\sigma} \hat{\bar{n}}_{r+1,\bar{\sigma}}  \hat{\bar{n}}_{r-1,\bar{\sigma}} \hc_{r-1,\sigma}^\dagger 
+\sum_{\sigma} \hat{\bar{n}}_{r,\bar{\sigma}} \hc^\dagger_{r+1,\sigma} \hn_{r+1,\bar{\sigma}}  \hn_{r-1,\bar{\sigma}} \hc_{r-1,\sigma}$.
Then we have 
 \eqq{
 \hU\hV_1\hU^\dagger|{\bf r}\rangle|\bsig\rangle|\boldeta\rangle  = \hU\hV_1\Bigl(\prod_{i=1}^{N_s} \hc^\dagger_{r_i,\sigma_i}\Bigl) \Bigl(\prod_{j=1}^{N_{\eta}} \ha^\dagger_{\bar{r}_j,\eta_j}\Bigl)|{\rm vac}\rangle 
= \sum_i  \hU \hV_1(r_i)\hU^\dagger|{\bf r}\rangle|\bsig\rangle|\boldeta\rangle + \sum_j  \hU \hV_1(\bar{r}_j)\hU^\dagger|{\bf r}\rangle|\bsig\rangle|\boldeta\rangle.
 }
 The first term has the property $\hU \hV_1(r_i)\hU^\dagger|{\bf r}\rangle|\bsig\rangle|\boldeta\rangle \propto  \delta(r_i +1 = r_{i+1}, r_i-1 \neq r_{i-1}) $.
 In this case, we can show that
\eqq{
& \sum_\sigma \hn_{r_i,\bar{\sigma}} \hc_{r_i+1,\sigma} \hat{\bar{n}}_{r_i+1,\bar{\sigma}}  \hat{\bar{n}}_{r_i-1,\bar{\sigma}} \hc_{r_i-1,\sigma}^\dagger  
(\cdots \hc^\dagger_{r_i+1,\sigma_{i+1}}\hc^\dagger_{r_i,\sigma_i} a^\dagger_{r_i-1,\eta'}\cdots)|{\rm vac}\rangle \\\nonumber
&=\delta_{\eta',\downarrow} \delta_{\bar{\sigma}_i,\sigma_{i+1}} (\cdots \ha^\dagger_{r_i+1,\eta'}\hc^\dagger_{r_i,\sigma_i}\hc^\dagger_{r_i-1,\sigma_{i+1}}\cdots)|{\rm vac}\rangle
}
and 
\eqq{
&\sum_\sigma \hat{\bar{n}}_{r_i,\bar{\sigma}} \hc^\dagger_{r_i+1,\sigma} \hn_{r_i+1,\bar{\sigma}}  \hn_{r_i-1,\bar{\sigma}} \hc_{r_i-1,\sigma}
(\cdots \hc^\dagger_{r_i+1,\sigma_{i+1}}\hc^\dagger_{r_i,\sigma_i} a^\dagger_{r_i-1,\eta'}\cdots)|{\rm vac}\rangle \\\nonumber
&=\delta_{\eta',\uparrow} \delta_{\bar{\sigma}_i,\sigma_{i+1}} (\cdots \ha^\dagger_{r_i+1,\eta'}\hc^\dagger_{r_i,\sigma_i}\hc^\dagger_{r_i-1,\sigma_{i+1}}\cdots)|{\rm vac}\rangle.
}
Thus, we have 
\eqq{
\hV_1(r_i)\hU^\dagger|{\bf r}\rangle|\bsig\rangle|\boldeta\rangle =  \delta(r_i +1 = r_{i+1}, r_i-1 \neq r_{i-1}) \delta_{\bar{\sigma}_i,\sigma_{i+1}} (\cdots \ha^\dagger_{r_i+1,\eta'}\hc^\dagger_{r_i,\sigma_i}\hc^\dagger_{r_i-1,\sigma_{i+1}}\cdots)|{\rm vac}\rangle
}
and
\eqq{
\sum_i \hU \hV_1(r_i)\hU^\dagger|{\bf r}\rangle|\bsig\rangle|\boldeta\rangle = \sum_i (-\hc^\dagger_{r_i-1}\hc_{r_i+1}\hn_{r_i})|{\bf r}\rangle\otimes (\hs^+_{i+1}\hs^-_i + \hs^-_{i+1}\hs^+_i)|\bsig\rangle \otimes|\boldeta\rangle.
}

The second term has the property $\hU \hV_1(\bar{r}_j)\hU^\dagger|{\bf r}\rangle|\bsig\rangle|\boldeta\rangle \propto  \delta(\bar{r}_j +1 \neq \bar{r}_{j+1}, \bar{r}_j-1 = \bar{r}_{j-1})$ and we can evaluate it in the same way as the first term. Then, we obtain 
\eqq{
\hV_1(\bar{r}_j)\hU^\dagger|{\bf r}\rangle|\bsig\rangle|\boldeta\rangle =- \delta(\bar{r}_j +1 \neq \bar{r}_{j+1}, \bar{r}_j-1 = \bar{r}_{j-1}) \delta_{\bar{\eta}_j,\eta_{j-1}} 
(\cdots \ha^\dagger_{\bar{r}_j+1,\eta_{j-1}}\ha^\dagger_{\bar{r}_j,\eta_j}\hc^\dagger_{\bar{r}_j-1,\sigma'}\cdots)|{\rm vac}\rangle
}
and
\eqq{
\sum_j \hU \hV_1(\bar{r}_j)\hU^\dagger|{\bf r}\rangle|\bsig\rangle|\boldeta\rangle = \sum_j (-\hat{\bar{n}}_{\bar{r}_j}\hc^\dagger_{\bar{r}_j-1}\hc_{\bar{r}_j+1})|{\bf r}\rangle\otimes |\bsig\rangle \otimes (\heta^+_{j}\heta^-_{j-1} + \heta^-_{j}\heta^+_{j-1}) |\boldeta\rangle.
}

From these results we can derive 
\eqq{
&\langle\boldeta'|\langle \bsig' |\langle \Psi^{\rm GS}_{\rm SF} | \hU \hV_{1} \hU^\dagger|\Psi^{\rm GS}_{\rm SF}\rangle|\bsig\rangle|\boldeta\rangle 
\\&=
 -\delta_{\boldeta',\boldeta} \sum_i \langle \Psi^{\rm GS}_{\rm SF} | \sum_{{\bf r}\in \Lambda}\psi_{\rm c}({\bf r}) \hc^\dagger_{r_i-1}\hc_{r_i+1}\hn_{r_i} |{\bf r}\rangle \langle \bsig' | \hs^+_{i+1} \hs^-_{i}+ \hs^-_{i+1} \hs^+_{i} | \bsig \rangle 
\\&\;\;\,\,\, - \delta_{\bsig',\bsig} \sum_j  \langle \Psi^{\rm GS}_{\rm SF} | \sum_{{\bf r}\in \Lambda}\psi_{\rm c}({\bf r}) \hat{\bar{n}}_{\bar{r}_j}\hc^\dagger_{\bar{r}_j-1}\hc_{\bar{r}_j+1} |{\bf r}\rangle \langle \boldeta' | \heta^+_{j} \heta^-_{j-1}+\heta^-_{j} \heta^+_{j-1} | \boldeta \rangle.
 }
 In a large enough system, we have 
 $\langle \Psi^{\rm GS}_{\rm SF} | \sum_{{\bf r}\in \Lambda}\psi_{\rm c}({\bf r}) \hc^\dagger_{r_i-1}\hc_{r_i+1}\hn_{r_i} |{\bf r}\rangle \simeq $$ \frac{1}{N_s}  \sum_i \langle \Psi^{\rm GS}_{\rm SF} | \sum_{{\bf r}\in \Lambda}\psi_{\rm c}({\bf r}) \hc^\dagger_{r_i-1}\hc_{r_i+1}\hn_{r_i} |{\bf r}\rangle =  \frac{1}{N_s} \sum_r \langle \Psi^{\rm GS}_{\rm SF} | \sum_{{\bf r}\in \Lambda}\psi_{\rm c}({\bf r}) \hc^\dagger_{r-1}\hc_{r+1}\hn_{r} |{\bf r}\rangle$ and $ \langle \Psi^{\rm GS}_{\rm SF} | \sum_{{\bf r}\in \Lambda}\psi_{\rm c}({\bf r}) \hat{\bar{n}}_{\bar{r}_j}\hc^\dagger_{\bar{r}_j-1}\hc_{\bar{r}_j+1} |{\bf r}\rangle \simeq \frac{1}{N_\eta} \sum_j  \langle \Psi^{\rm GS}_{\rm SF} | \sum_{{\bf r}\in \Lambda}\psi_{\rm c}({\bf r}) \hat{\bar{n}}_{\bar{r}_j}\hc^\dagger_{\bar{r}_j-1}\hc_{\bar{r}_j+1} |{\bf r}\rangle = \frac{1}{N_\eta}\sum_r  \langle \Psi^{\rm GS}_{\rm SF} | \sum_{{\bf r}\in \Lambda}\psi_{\rm c}({\bf r}) \hat{\bar{n}}_{r}\hc^\dagger_{r-1}\hc_{r+1} |{\bf r}\rangle$. Thus, we get Eq.~\eqref{eq:V1}.

  One can evaluate  $\tilde{x},\tilde{y},\tilde{x}',\tilde{y}'$ for $|\Psi^{\rm GS}_{\rm SF}\rangle$ using the Wick's theorem and $ \langle \hc^\dagger_l \hc_m \rangle = \frac{1}{\pi} \frac{\sin(\pi n_s (l-m))}{l-m}$ in the thermodynamic limit.
  When we combine the conclusions for $\hH_{\rm 2-site}$ and $\hH_{\rm 3-site}$, we obtain the expression shown in the main text.

\section{Correlation functions}
In this section, we discuss how to evaluate the spin correlations ($\chi_{s,a} (r) \equiv \langle \hs^a(r) \hs^a(0)\rangle$) and $\eta$-spin correlations ($\chi_{\eta,a} (r)\equiv \langle \heta^a(r) \heta^a(0)\rangle$) in detail. Here, $a=x,y,z$. Given the exact form of the wave function (Eq.~(5) in the main text),  
we have 
\eqq{
 \chi_{s,a}(r) = \sum_{m=2}^{r+1} Q^r_{\rm SF}(m) \chi^{\rm (SQ)}_{s,a} (m-1),\;\;\; \chi_{\eta,a}(r) = \sum_{m=2}^{r+1} \bar{Q}^r_{\rm SF}(m) \chi^{\rm (SQ)}_{\eta,a} (m-1). \label{eq:correlations}
}

Here  $Q^r_{\rm SF}(m) = \langle \hn_0 \hn_r \delta(\sum_{l=0}^r \hn_l -m) \rangle_{\rm SF} $ and $\bar{Q}^r_{\rm SF}(m) = \langle \hat{\bar{n}}_0 \hat{\bar{n}}_r \delta(\sum_{l=0}^r \hat{\bar{n}}_l -m) \rangle_{\rm SF} $,
where $\hat{\bar{n}} = 1-\hn$ and both expectation values are calculated with $|\Psi^{\rm GS}_{\rm SF}\rangle$.
The former  is the probability that the system has $m$ singlons (singly occupied sites) in $[0,r]$,
and the latter  is the probability that the system has $m$ doublons or holons in $[0,r]$.
$\chi^{\rm (SQ)}_{s,a}(m) = \langle \hs^a(m) \hs^a(0) \rangle_{\rm spin,squeezed}$
 is the correlation function in the squeezed spin space.
$\chi^{\rm (SQ)}_{\eta,a}(m) = \langle \heta^a(m) \heta^a(0) \rangle_{\rm \eta-spin,squeezed}$
 is the correlation function in the squeezed $\eta$-spin space.
 As is emphasized in the main text, the spin correlations are the same as those for the Ogata-Shiba states in equilibrium, hence the discussions in the previous works \cite{Pruschke1991PRB,Sorella1990PRL}
 are directly applicable. In the case of the $\eta$-spin correlations, we need to extend the previous discussions.
 
\subsection{Numerical analysis}
In order to evaluate $\chi_{\eta,a}(r)$ ($\chi_{s,a}(r)$) numerically, we separately calculate $\bar{Q}^r_{\rm SF}$ ($Q^r_{\rm SF}$) and $\chi^{\rm (SQ)}_{\eta,a}$ ($\chi^{\rm (SQ)}_{s,a}$) in the thermodynamic limit. The $\eta$-spin and spin correlation functions in their squeezed space can be obtained in the thermodynamic limit by simulating the XXZ model with the infinite time-evolving block decimation (iTEBD)~\cite{Vidal2003PRL}.
 $\bar{Q}^r_{\rm SF}(m)$ and $Q^r_{\rm SF}(m)$ can be obtained by considering the Fourier components of these functions. 
 To simplify the notation in this section, we introduce $\bar{P}^{r}_{\rm SF}(m)= \langle \hat{\bar{n}}_0 \hat{\bar{n}}_r \delta(\sum_{l=1}^{r-1} \hat{\bar{n}}_l -m) \rangle_{\rm SF}$ $(= \bar{Q}^r_{\rm SF}(m+2))$ and $P^{r}_{\rm SF}(m)=\langle \hn_0 \hn_r \delta(\sum_{l=1}^{r-1} \hn_l -m) \rangle_{\rm SF}$ $(= Q^r_{\rm SF}(m+2))$.
Note that these function are zero for $m \geq r$.
 Thus, we can assume that these functions are defined only for $0\leq m < N'_{r}$ ($N'_{r}\geq r$ is an arbitrary number), and consider the Fourier components
 \eqq{
 P^{r}_{\rm SF}(k) \equiv \sum_{m=0}^{N_r'-1} e^{-ikm} P^r_{\rm SF}(m) = \Big\langle \hn_r \hn_0 \prod_{l=1}^{r-1} e^{-ik\hn_l}\Big\rangle_{\rm SF},\;\;
 \bar{P}^{r}_{\rm SF}(k) \equiv \sum_{m=0}^{N_r'-1} e^{-ikm} \bar{P}^r_{\rm SF}(m) = \Big\langle \hat{\bar{n}}_r \hat{\bar{n}}_0 \prod_{l=1}^{r-1} e^{-ik\hat{\bar{n}}_l}\Big\rangle_{\rm SF},
 }
 where $k=2\pi j/N_r'$ with $j\in[0,N_r')$.
 Using $ \hn_r = \frac{1}{e^{-ik} -1} (e^{-ik\hn_r}-1)$ and $ \hat{\bar{n}}_r = \frac{1}{e^{-ik} -1} (e^{-ik\hat{\bar{n}}_r}-1)$, these functions can be expressed as 
  \eqq{
 P^{r}_{\rm SF}(k)
 &= \frac{1}{(e^{-ik} -1)^2} \Bigl[ \Big\langle \exp\Big[-ik\sum_{l=0}^r \hn_l \Big] + \exp\Big[-ik\sum_{l=0}^{r-2} \hn_l \Big] - 2\exp\Big[-ik\sum_{l=0}^{r-1} \hn_l \Big]  \Big\rangle_{\rm SF}\Bigl],\\
 \bar{P}^{r}_{\rm SF}(k)
 & = \frac{1}{(e^{-ik} -1)^2} \Bigl[ \Big\langle \exp\Big[-ik\sum_{l=0}^r \hat{\bar{n}}_l \Big] + \exp\Big[-ik\sum_{l=0}^{r-2} \hat{\bar{n}}_l \Big] - 2\exp\Big[-ik\sum_{l=0}^{r-1} \hat{\bar{n}}_l \Big]  \Big\rangle_{\rm SF}\Big].
 }
 The right hand side can be evaluated using the following identities:
 \begin{enumerate}
 \item \eqq{\Big\langle \exp\Big[-ik\sum_{l=0}^r \hn_l \Big] \Big\rangle_{\rm SF} =  \Big\langle \prod_{l=0}^{r} (1- (1-e^{-ik}) n_l)\Big\rangle_{\rm SF} 
 ={\rm det} [\delta_{nm}-\langle \hc_m^\dagger \hc_n\rangle_{\rm SF} (1-e^{-ik})]^r_{n,m=0}, \label{eq:exp_n} \\
  \Big\langle \exp\Big[-ik\sum_{l=0}^r \hat{\bar{n}}_l \Big] \Big\rangle_{\rm SF} =  \Big\langle \exp[-ik\sum_{l=0}^r (1-\hat{n}_l) ] \Big\rangle_{\rm SF}  
 =e^{-ik (r+1)}{\rm det} [\delta_{nm}-\langle \hc_m^\dagger \hc_n\rangle_{\rm SF} (1-e^{ik})]^r_{n,m=0}. 
 }
 
 \item In the thermodynamic limit, we have 
 \eqq{
 \langle \hc^\dagger_l \hc_m \rangle = \frac{1}{\pi} \frac{\sin(\pi n_s (l-m))}{l-m}.
 }
 
 \end{enumerate}
Finally, taking the inverse Fourier transform of  $P^{r}_{\rm SF}(k)$ and $ \bar{P}^{r}_{\rm SF}(k)$, we can evaluate  $Q^r_{\rm SF}(m)$ and $\bar{Q}^r_{\rm SF}(m)$ in the thermodynamic limit. 
We note that a previous work~\cite{Pruschke1991PRB} evaluated the correlation functions  $\chi_{s,a}(r)$ using their expressions in Fourier space, which limits the analysis to finite size systems.

\begin{figure}[t]
  \centering
    \hspace{-0.cm}
    \vspace{0.0cm}
\includegraphics[width=170mm]{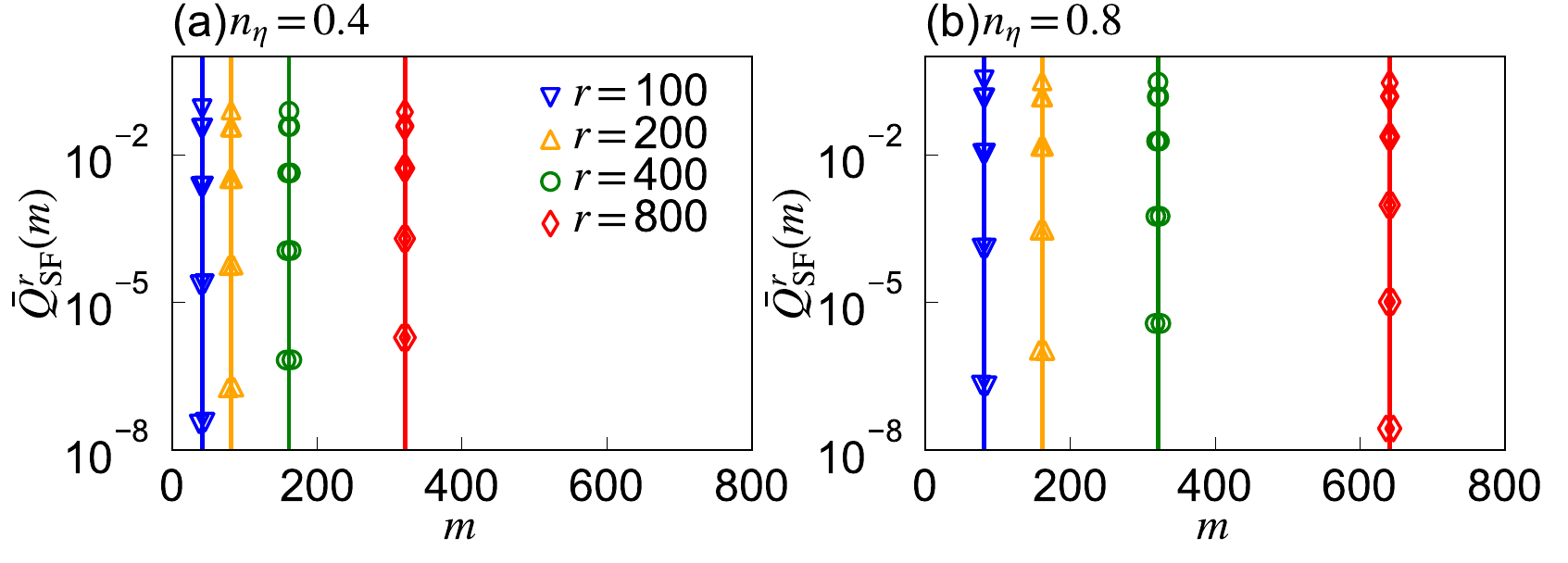} 
  \caption{Numerically evaluated $\bar{Q}^r_{\rm SF}(m)$ for the indicated values of $r$ and $n_\eta$. The vertical lines indicate $m=r\;n_\eta +1$. Since $\bar{Q}^r_{\rm SF}(m)$ quickly approaches zero away from $m=r\;n_\eta +1$, we use the logarithmic scale for the $y$ axis. Although the data points appear to lie on the lines of $m=r\;n_\eta +1$, they correspond to different values of $m$.}
  \label{fig:Q_bar}
\end{figure}

Before we end this section, we show the numerical results for $\bar{Q}^r_{\rm SF}(m)$ in Fig.~\ref{fig:Q_bar}.
As can be anticipated from its moments (discussed in the next section), the function is sharply peaked at $m=r\; n_\eta+1$.

\subsection{Analytic considerations}
One can also analytically evaluate the asymptotic behavior of the spin and $\eta$-spin correlation functions by extending the analysis of the spin correlation functions for the equilibrium Hubbard model~\cite{Sorella1990PRL}.
The analytic expressions for the asymptotic behavior of the correlation functions of the Heisenberg model and the XXZ model are well known, and these functions correspond to $\chi^{\rm (SQ)}_{s}$ and  $\chi^{\rm (SQ)}_{\eta}$, respectively. Specifically, we have $\chi^{\rm (SQ)}_{s}(m)= (-1)^m f_s(m)$ with $f_s(m)\rightarrow A \ln^\frac{1}{2} (m)/m$.
$\chi^{\rm (SQ)}_{\eta}$ corresponds to the correlations of the one-dimensional XXZ model
\eqq{
\hH = \sum_i -J_X(\heta_i^x \heta_{i+1}^x + \heta_i^y \heta_{i+1}^y) + J_Z \heta_i^z \heta_{i+1}^z
}
with $J_X>0$. 
The spin-spin correlation functions for this model behave as~\cite{Giamarchi_book,Lukyanov99PRB,Lukyanov03NPB} 
\begin{align}
 \braket{\heta_{r}^{x}\heta_{0}^{x}}
   =&A_{0}r^{-\alpha}-A_{1}(-1)^{r}r^{-\alpha-\frac{1}{\alpha}}
     +\cdots ,
 \\
 \braket{\heta_{r}^{z}\heta_{0}^{z}}
   =&-\frac{1}{4\pi^{2}\alpha} r^{-2}+
     (-1)^{r} B_{1} r^{-\frac{1}{\alpha}}+\cdots,
\end{align}
where $\alpha\equiv 1-\frac{1}{\pi}\arccos(\Delta)$ and $\Delta = J_Z/J_X$.

Depending on the component of the correlation function and the system parameters, the asymptotic behavior can be expressed with an (asymptotically) smooth function $f_\eta(m)$ as 
(A) $\chi^{\rm(SQ)}_{\eta} (m) = (-)^mf_\eta(m)$ or (B) $\chi^{\rm (SQ)}_{\eta} (m) = f_\eta(m)$.
In the following, we simply use $f(m)$ to express either of $f_s(m)$ or $f_\eta(m)$.
We can easily prove the following statement about $f(m)$:
\begin{itembox}[l]{Properties of $f(m)$}
\begin{enumerate}[(1)]
\item We assume $f(m) \rightarrow \Gamma \frac{\ln^\sigma(m)}{m}$ ($\sigma$ is positive). For a given (but large enough) $r$, we consider $m>m'\geq r$.
We have 
\eqq{
\Bigl| \frac{f(m)-f(m')}{m-m'} \Bigl| \leq M(r) = \frac{2\Gamma}{r^2} \ln^\sigma(r). \label{eq:f_prop_1}
}

\item We assume $f(m) \rightarrow \Gamma m^{-\gamma}$ ($\gamma\geq0$). For a given (but large enough) $r$, we consider $m>m'\geq r$.
We have 
\eqq{
\Bigl| \frac{f(m)-f(m')}{m-m'} \Bigl| \leq M(r) = \frac{C}{r^{\gamma +1}} .  \label{eq:f_prop_2}
}
\end{enumerate}
  \end{itembox}
  Here, $M(r)$ represents a function that sets the upper-bound.

In Ref.~\cite{Sorella1990PRL},  the moments of $Q^r_{\rm SF}(m)$ are evaluated. The following properties hold:
\begin{itembox}[l]{Properties of $Q^r_{\rm SF}(m)$}
Assuming $L\rightarrow \infty$ and $N_s \gg r$, we have the following asymptotic behaviors:
\begin{enumerate}[(i)]
\item \eqq{
Z\equiv \sum_{m=2}^{r+1} Q^{r}_{\rm SF}(m) = n_s^2 + \mathcal{O}(1/r^2).
}
\item The moments of $Q^{r}_{\rm SF}$ asymptotically behave (in leading order) as 
\eqq{
\langle m \rangle &= Z^{-1}\sum_{m=2}^{r+1} m Q^{r}_{\rm SF}(m)  = r n_s +1, \\
\langle m^2 \rangle &= Z^{-1}\sum_{m=2}^{r+1} m^2 Q^{r}_{\rm SF}(m)  = \langle m\rangle^2 + \frac{1}{\pi^2} \ln r.
}
\end{enumerate}
  \end{itembox}
  
 Using these properties of $Q^r_{\rm SF}(m)$, one can easily show the following properties of $\bar{Q}^r_{\rm SF}(m)$:
\begin{itembox}[l]{Properties of $\bar{Q}^r_{\rm SF}(m)$}
Assuming $L\rightarrow \infty$ and $N_{\eta} \gg r$, we have the following asymptotic behaviors:
\begin{enumerate}[(i)]
\item \eqq{
\bar{Z}\equiv \sum_{m=2}^{r+1} \bar{Q}^{r}_{\rm SF}(m) = n_{\eta}^2 + \mathcal{O}(1/r^2).
}
\item The moments of $\bar{Q}^{r}_\text{SF}$ asymptotically behave (in leading order) as 
\eqq{
\overline{\langle m \rangle} &\equiv \bar{Z}^{-1}\sum_{m=2}^{r+1} m \bar{Q}^{r}_\text{SF}(m)  = r n_{\eta} +1, \\
\overline{\langle m^2 \rangle} &\equiv \bar{Z}^{-1}\sum_{m=2}^{r+1} m^2 \bar{Q}^{r}_\text{SF}(m)  = \overline{\langle m\rangle}^2 + \frac{1}{\pi^2} \ln r.
}
\end{enumerate}
  \end{itembox}
  
The first statement can be verified easily from $\langle\hat{\bar{n}}_r \hat{\bar{n}}_0 \rangle  = \langle \hat{\bar{n}}_r \rangle \langle \hat{\bar{n}}_0 \rangle + \langle \hc_r \hc_0^\dagger \rangle \langle \hc^\dagger_r \hc_0 \rangle$.
The second statement naturally follows from the corresponding properties of $Q^r_\text{SF}$.
The Hamiltonian of the spinless fermions is $\hH_{\rm SF,0} = -t_{\rm hop} \sum_r [\hc^\dagger_r \hc_{r+1} + \hc^\dagger_{r+1} \hc_r]$, and we denote the ground state for the system with $L$ sites and $N_s$ particles by $|\Psi_{\rm SF}(N_s,L)\rangle$. This Hamiltonian is invariant under the particle-hole transformation $\hU_{\rm PH}$, which maps $\hc^\dagger_r \rightarrow (-)^r \hc_r$, while $|\Psi_{\rm SF}(N_s,L)\rangle$ is transformed to $|\Psi_{\rm SF}(N_\eta,L)\rangle$. 
Thus, we have 
\eqq{  
\bar{Q}^r_{\rm SF}(m; N_s, L) &= \langle  \Psi_{\rm SF}(N_s,L) | \hU^\dagger_{\rm PH} \hU_{\rm PH} \hat{\bar{n}}_0 \hat{\bar{n}}_r \delta\Big(\sum_{l=0}^r \hat{\bar{n}}_l -m\Big)  \hU^\dagger_{\rm PH} \hU_{\rm PH}| \Psi_{\rm SF}(N_s,L)\rangle \\ \nonumber
&= \langle  \Psi_{\rm SF}(N_{\eta},L) | \hn_0 \hn_r \delta\Big(\sum_{l=0}^r \hn_l -m\Big) | \Psi_{\rm SF}(N_{\eta},L)\rangle = Q^r_{\rm SF}(m;N_{\eta}, L).
}
We can now reach the conclusions. 
We note that the above properties indicate that  the weight of $\bar{Q}^r_{\rm SF}(m)$ is concentrated around $ \overline{\langle m \rangle}$, which has indeed been numerically verified in the previous section.
Therefore, one can naively expect that in Eq.~\eqref{eq:correlations} one may use $f(\overline{\langle m \rangle})$ to estimate the contribution from  $\chi_\eta^{\rm (SQ)}(m)$.

With this, the following statements hold:
 \begin{itembox}[l]{Case (A): $\chi^{\rm (SQ)}_{\eta}(m) = (-)^m f(m)$}
\begin{enumerate}
\item If $f(m) \rightarrow \Gamma \frac{\ln^\sigma(m)}{m}$, for large enough $r$,
\eqq{
\Bigl[ \sum_{m=2}^{r+1} \bar{Q}^r_{\rm SF}(m) (-)^m f(m) \Bigl] - \Bigl[ \Big\{ \sum_{m=2}^{r+1} \bar{Q}^r_{\rm SF}(m) (-)^m\Big\} f(\overline{\langle m\rangle} ) \Bigl]  \leq \mathcal{O}\Big(\frac{\ln^{|\sigma|+1}(r)}{r^2}\Big).
}

\item If $f(m) \rightarrow \Gamma m^{-\gamma}$, for large enough $r$,
\begin{enumerate}
\item \underline{$\gamma\geq 1$}
\eqq{
\Bigl[ \sum_{m=2}^{r+1} \bar{Q}^r_{\rm SF}(m) (-)^m f(m) \Bigl] - \Bigl[ \Big\{ \sum_{m=2}^{r+1} \bar{Q}^r_{\rm SF}(m) (-)^m\Big\} f(\overline{\langle m\rangle}) \Bigl]  \leq \mathcal{O}\Big(\frac{\ln^{1}(r)}{r^2}\Big),
}
\item \underline{$\gamma<1$}
\eqq{
\Bigl[ \sum_{m=2}^{r+1} \bar{Q}^r_{\rm SF}(m) (-)^m f(m) \Bigl] - \Bigl[ \Big\{ \sum_{m=2}^{r+1} \bar{Q}^r_{\rm SF}(m) (-)^m\Big\} f(\overline{\langle m\rangle} ) \Bigl]  \leq \mathcal{O}\Big(\frac{\ln^{\frac{1}{2}}(r)}{r^{1+\gamma}}\Big).
}
\end{enumerate}
\end{enumerate}
\end{itembox}

 \begin{itembox}[l]{Case (B): $\chi^{\rm (SQ)}_{\eta}(m) = f(m)$}
\begin{enumerate}
\item  If $f(m) \rightarrow \Gamma \frac{\ln^\sigma(m)}{m}$, for large enough $r$,
\eqq{
\Bigl[ \sum_{m=2}^{r+1} \bar{Q}^r_{\rm SF}(m) f(m) \Bigl] - \Bigl[ \Big\{ \sum_{m=2}^{r+1} \bar{Q}^r_{\rm SF}(m)\Big\} f(\overline{\langle m\rangle} ) \Bigl]  \leq \mathcal{O}\Big(\frac{\ln^{|\sigma|+1}(r)}{r^2}\Big).
}

\item If $f(m) \rightarrow \Gamma m^{-\gamma}$, for large enough $r$,
\begin{enumerate}
\item \underline{$\gamma\geq 1$}
\eqq{
\Bigl[ \sum_{m=2}^{r+1} \bar{Q}^r_{\rm SF}(m) f(m) \Bigl] - \Bigl[ \Big\{ \sum_{m=2}^{r+1} \bar{Q}^r_{\rm SF}(m) \Big\} f(\overline{\langle m\rangle} ) \Bigl]  \leq \mathcal{O}\Big(\frac{\ln^{1}(r)}{r^2}\Big),
}
\item \underline{$\gamma<1$}
\eqq{
\Bigl[ \sum_{m=2}^{r+1} \bar{Q}^r_{\rm SF}(m)  f(m) \Bigl] - \Bigl[ \Big\{ \sum_{m=2}^{r+1} \bar{Q}^r_{\rm SF}(m) \Big\} f(\overline{\langle m\rangle} ) \Bigl]  \leq \mathcal{O}\Big(\frac{\ln^{\frac{1}{2}}(r)}{r^{1+\gamma}}\Big).
}
\end{enumerate}
\end{enumerate}
\end{itembox}

These properties can be demonstrated following the arguments presented for the spin correlations in Ref.~[\onlinecite{Sorella1990PRL}].
Since the proofs for the case (A) and the case (B) are almost identical, we focus on the case (A).
We split the left hand side into $R=R_1 + R_2$, where
\eqq{
R_1 =  \sum_{m=2}^{\overline{\langle  m \rangle}/2} \bar{Q}^r_{\rm SF}(m) (-)^m [f(m) - f(\overline{\langle m\rangle} )], \;\;\;\;
R_2 =  \sum_{m=\frac{\overline{\langle m \rangle}}{2}+1}^{r+1} \bar{Q}^r_{\rm SF}(m) (-)^m [f(m) - f(\overline{\langle m\rangle})].
}
We also set $A$ to be an upper bound of $|f(m)|$. First, we consider the behavior of $R_1$.
Since $\bar{Q}^r_{\rm SF}(m)$ is positive definite, 
we have 
\eqq{
R_1\leq \sum_{m=2}^{\overline{\langle m \rangle}/2} \bar{Q}^r_{\rm SF}(m)[ |f(m)| + |f(\overline{\langle m\rangle})| ] \leq  2A \sum_{m=2}^{\overline{\langle m \rangle}/2} \bar{Q}^r_{\rm SF}(m). 
}
Considering  
\eqq{
\frac{\overline{\langle m \rangle}^2}{4} \sum_{m=2}^{\overline{\langle m \rangle}/2} \bar{Q}^r_{\rm SF}(m) \leq  \sum_{m=2}^{r+1} Q^r_{\rm SF}(m) (m-\overline{\langle m \rangle})^2 = \overline{\langle m^2\rangle} - \overline{\langle m\rangle}^2 = \frac{1}{\pi^2} \ln r,
}
we have 
\eqq{
R_1 \leq \mathcal{O}\Bigl(\frac{\ln r}{r^2}\Bigl).
}

As for $R_2$, we have 
\eqq{
R_2  \leq M\Bigl( \frac{\overline{\langle m \rangle}}{2}\Bigl)  \sum_{m = \frac{\overline{\langle m \rangle}}{2} + 1}^{r+1} \bar{Q}^r_{\rm SF}(m) | m - \overline{\langle m\rangle}| \nonumber 
 \leq M\Bigl( \frac{ \overline{\langle m \rangle}}{2}\Bigl)  \sum_{m = 2}^{r+1} \bar{Q}^r_{\rm SF}(m) | m - \overline{\langle m\rangle}|.
}
Remember that $M(r)$ has been introduced in Eqs.~\eqref{eq:f_prop_1} and \eqref{eq:f_prop_2}.
Regarding the summation as the inner product of $\{\bar{Q}^r_{\rm SF}(m)\}^{\frac{1}{2}} $ and $\{\bar{Q}^r_{\rm SF}(m)\}^{\frac{1}{2}} | m - \overline{\langle m\rangle}|$, we use the Schwarz inequality.
This yields 
\eqq{
R_2 \leq M\Bigl( \frac{\overline{\langle m \rangle}}{2}\Bigl) \bar{Z}^\frac{1}{2} \Bigl( \sum_{m = 2}^{r+1}  | m - \overline{\langle m\rangle}|^2 \bar{Q}^r_{\rm SF}(m) \Bigl)^\frac{1}{2} = M\Bigl( \frac{\overline{\langle m \rangle}}{2}\Bigl) \mathcal{O}(\ln^\frac{1}{2} r ).
}
Therefore, the behavior of $R_2$ depends on the behavior of $f(m)$ as follows:
\begin{enumerate}
\item If $f(m) \rightarrow \Gamma \frac{\ln^\sigma(m)}{m}$, we have 
\eqq{
R_2 \leq \mathcal{O}\Big(\frac{\ln^\sigma r}{r^2} \Big) \mathcal{O}\Big(\ln^\frac{1}{2}r\Big) = \mathcal{O}\Big(\frac{\ln^{\sigma + \frac{1}{2}} r}{r^2}\Big).
}

\item If $f(m)\rightarrow \Gamma m^{-\gamma}$, we have 
\eqq{
R_2 \leq \mathcal{O}\Big(\frac{1}{r^{1+\gamma}}\Big) \mathcal{O}\Big(\ln^\frac{1}{2}r\Big) = \mathcal{O}\Big(\frac{\ln^{\frac{1}{2}} r}{r^{1+\gamma}}\Big).
}
\end{enumerate}
Combining the behavior of $R_1$ and $R_2$, we obtain the conclusion.

As for the case (A), following the discussion for $\sum_{m=2}^{r+1} Q^r_{\rm SF}(m) (-)^m$ in the previous works~\cite{Sorella1990PRL,Holger1990PRB}, we can show that
\eqq{
\sum_{m=2}^{r+1} \bar{Q}^r_{\rm SF}(m) (-)^m \propto \frac{\cos(\pi n_\eta r)}{r^{\frac{1}{2}}}
} 
for large enough $r$.
From the above mentioned properties of the case (A), we can conclude that the followings asymptotic behavior holds:

\begin{itembox}[l]{Case (A): When $\chi^{\rm (SQ)}_{\eta}(m) = (-)^mf(m)$}
\begin{enumerate}
\item If $f(m) \rightarrow \Gamma \frac{\ln^\sigma(m)}{m}$, we have the asymptotic form 
\eqq{
\chi_\eta(r)  \propto \cos(\pi n_\eta r) \frac{\ln^\sigma(r)}{r^\frac{3}{2}}.
}

\item If $f(m)\rightarrow \Gamma m^{-\gamma}$, we have the asymptotic form  
\begin{enumerate}
\item \underline{$\gamma<\frac{3}{2}$}
\eqq{
\chi_\eta(r) \propto \cos(\pi n_\eta r) \frac{1}{r^{\frac{1}{2}+\gamma}},
}
\item \underline{$\gamma\geq\frac{3}{2}$}
\eqq{
 \chi_\eta(r)  \leq \mathcal{O}\Big(\frac{\ln r}{r^2}\Big).
}
\end{enumerate}
\end{enumerate}
\end{itembox}

These expressions show that for the oscillatory component, the contribution from the spinless fermions provides an additional factor $r^{-1/2}$, so that the correlation decays faster than in the squeezed space.

As for case (B), since $\sum_{m=2}^{r+1} \bar{Q}^r_{\rm SF}(m)$ approaches a constant, we can conclude that the following asymptotic behavior holds:
\begin{itembox}[l]{Case (B): When $\chi^{\rm (SQ)}_{\eta}(m) = f(m)$}
\begin{enumerate}
\item If $f(m) \rightarrow \Gamma \frac{\ln^\sigma(m)}{m}$, we have the asymptotic form 
\eqq{
\chi_\eta(r) \propto \frac{\ln^\sigma(r)}{r}.
}

\item If $f(m)\rightarrow \Gamma m^{-\gamma}$, we have the asymptotic form 
\begin{enumerate}
\item  \underline{$\gamma<2$}
\eqq{
\chi_\eta(r)  \propto \frac{1}{r^{\gamma}},
}
\item \underline{$\gamma\geq 2$}
\eqq{
\chi_\eta(r)   \leq \mathcal{O}\Big(\frac{\ln r}{r^2}\Big).
}
\end{enumerate}
\end{enumerate}
\end{itembox}

These expressions show that  there is no correction from the spinless fermions and the exponent is the same as that in the squeezed space.

\section{Evaluation of the central charge from iTEBD}
In this work, we evaluate the central charge using Eq.~(8) in the main text, following Ref.~[\onlinecite{Pollmann2013PRB}].
In iTEBD, we express the wave function of the system as an infinite matrix product state (MPS) in the canonical form,
\eqq{
|\Psi\rangle = \sum_{\cdots, \alpha_r,j_r,\alpha_{r+1},j_{r+1}} \cdots \Gamma^{[r]}_{\alpha_{r-1},j_r,\alpha_r} \Lambda^{[r]}_{\alpha_r}  \Gamma^{[r+1]}_{\alpha_{r},j_{r+1},\alpha_{r+1}} \Lambda^{[r+1]}_{\alpha_{r+1}} \cdots |\cdots, j_r,j_{r+1}, \cdots\rangle.
}
Here, $\Gamma^{[r]}$ is a three-way tensor of dimension $D\times d \times D$ and $\Lambda^{[r]}$ is a positive, real, square diagonal matrix.
$D$ is the cut-off dimension, $d$ is the dimension of the local states, and $j$ indicates the local states.
Translational invariance is imposed in the MPS. In practice, we introduce the $A$,$B$ sublattice structure and set
$\Gamma^A = \Gamma^{[2n]}=\Gamma^{[2n+2]}=\cdots$,  
$\Gamma^B = \Gamma^{[2n+1]}=\Gamma^{[2n+3]}=\cdots$ and $\Lambda^A = \Lambda^{[2n]}=\Lambda^{[2n+2]}=\cdots$ and $\Lambda^B = \Lambda^{[2n+1]}=\Lambda^{[2n+3]}=\cdots$.
The ground state is evaluated by the imaginary-time evolution and when the ground state does not break the symmetry between the sublattices, the result converges to 
$\Gamma^A=\Gamma^B$ and  $\Lambda^A=\Lambda^B$.
The entanglement entropies for the $AB$ and $BA$ bonds are given by
\eqq{
S^{[AB]}_E = -\sum_\alpha ( \Lambda^{A}_\alpha)^2 \ln ( \Lambda^{A}_\alpha)^2,\;\;  S^{[BA]}_E = -\sum_\alpha ( \Lambda^{B}_\alpha)^2 \ln ( \Lambda^{B}_\alpha)^2,
}
respectively. The bond-averaged entanglement entropy is defined as $S_E = \frac{S^{[AB]}_E + S^{[BA]}_E}{2}$.
On the other hand, the correlation length of the MPS is evaluated from the transfer matrix $T$,
\eqq{
T^{[X]}_{\alpha,\alpha';\beta,\beta'} = \sum_j \Gamma^{[X]}_{\alpha,j,\beta} (\Gamma^{[X]}_{\alpha',j,\beta'})^* \Lambda^{[X]}_\beta \Lambda^{[X]}_{\beta'}.
}
Here $X=A,B$ and ``*" denotes complex conjugation. The indices of $T$, $\alpha,\alpha';\beta,\beta'$, indicate that we regard this object as a $(D^2)\times(D^2)$ matrix.
Taking account of a possible inequivalence of the $A,B$ sublattices, we introduce the transfer matrix for two sites as $T^{AB} = T^{[A]}T^{[B]}$. 
Then the correlation length of the MPS can be defined as 
\eqq{
\xi_D  = -\frac{1}{\ln |\epsilon_2|},
}
where $\epsilon_2$ is the second largest right eigenvalue of $T^{AB}$. 
Note that the largest eigenvalue should be one.
To obtain $\epsilon_2$, one can use the Implicit Restarted Arnoldi Method~\cite{ARPACK}. 

\end{document}